\documentclass[aps,pre,twocolumn]{revtex4}
\usepackage{graphicx}
\usepackage{amssymb,amsfonts,amsmath}
\pdfoutput=1

\newenvironment{mylisting}
{\begin{list}{}{\setlength{\leftmargin}{1em}}\item\scriptsize\bfseries}
{\end{list}}

\begin{document}

\title{Characterizing Complex Particle Morphologies Through Shape Matching: Descriptors, Applications, and Algorithms}


\author{Aaron S. Keys$^1$}
\author{Christopher R. Iacovella$^1$}
\author{Sharon C. Glotzer$^{1,2}$}
\affiliation{$^1$Department of Chemical Engineering and $^2$Department of Materials Science and Engineering \\University of Michigan, Ann Arbor, Michigan 48109-2136}

\date{\today}

\begin{abstract}
%
%

Many standard structural quantities, such as order parameters and correlation functions, exist for common condensed matter systems, such as spherical and rod-like particles.  However, these structural quantities are often insufficient for characterizing the unique and highly complex structures often encountered in the emerging field of nano and microscale self-assembly, or other disciplines involving complex structures such as computational biology.  Computer science algorithms known as ``shape matching'' methods pose a unique solution to this problem by providing robust metrics for quantifying the similarity between pairs of arbitrarily complex structures.  This pairwise matching operation, either implicitly or explicitly, lies at the heart of most standard structural characterization schemes for particle systems.  By substituting more robust ``shape descriptors'' into these schemes we extend their applicability to structures formed from more complex building blocks.  Here, we describe several structural characterization schemes and shape descriptors that can be used to obtain various types of structural information about particle systems.  We demonstrate the application of shape matching algorithms to a variety of example problems, for topics including local and global structure identification and classification, automated phase diagram mapping, and the construction of spatial and temporal correlation functions.  The methods are applicable to a wide range of systems, both simulated and experimental, provided particle positions are known or can be accurately imaged.

\end{abstract}


\maketitle


It has been long recognized that in condensed matter systems there exists a strong connection between thermodynamics and particle packing~\cite{flory, onsager, bernal}.  Additionally, the spatial arrangement of particles in a given phase determines its mechanical, chemical, electrical, and optical properties.  As a result, it is natural to attempt to gain insight into systems in general by characterizing and monitoring both global and local structure.  In standard condensed systems, this is typically achieved by constructing order parameters or correlation functions that are sensitive to the way particles are arranged.  Several order parameters and correlation functions have been contrived for standard classes of condensed matter, including, e.g., systems of  rod-like and spherical particles~\cite{larson, mermin1968, tenwolde96, cna, kamienchiral}.  Standard examples include the $\bar{P}_2$ nematic order parameter for rod-like liquid crystalline systems and the bond order parameters of Nelson and coworkers for detecting crystalline ordering in systems of spherical particles~\cite{halperin78, snr83}.  These types of standard metrics have found widespread use in both computational condensed matter physics as well as the colloidal sciences, where standard systems of rod-like or spherical particles are often studied.



In the emerging field of colloidal and nanoscale self-assembly, unique building blocks can form assembled morphologies that often deviate from those expected in traditional condensed matter systems~\cite{glotzer07, glotzer2005, kumacheva,desimone2006}. Beyond the self-assembly of spherical~\cite{storhoff, akcora} and rod-like~\cite{nie2007, kempa} particles ~\cite{kumacheva, glotzer07}, examples of assembled systems include ordered structures formed from polyhedrally shaped ~\cite{amir09, tang2006, zhang2007, zhang2003, twistedribbons, magcubes, dipolecube, bettina} or patterned particles~\cite{mohwald2005, zhang2004, zhenlidiamond, devries2007, jackson2004}, and phase-separated domains reminiscent of those formed by block copolymers and surfactants~\cite{glotzer2005, zhang2003, park, reister, arthi2008, waddon2002, rotello}.  In these systems too, system stability and properties are, in many cases, strongly linked to their global structure and local packing~\cite{kumacheva, daniel2004,collier, storhoff, kempa, reinhard}.  However, constructing general order parameters for assemblies of particles of complex shape and interaction anisotropy~\cite{glotzer07, kumacheva} is considerably more challenging than for traditional condensed systems, where the particle shapes and morphologies are comparatively much simpler.  As a result of the increased complexity and vast design space, there are few ``model problems'' in nanoscale self-assembly for which generally applicable order parameters can be defined.  This has lead many recent studies of assembled systems to rely heavily on visual inspection or \textit{ad hoc} analysis for characterizing structures, which are often more time consuming and less accurate than mathematical analysis.


In this article, we address the problem of creating general structural metrics for complex colloidal and nanoscale assemblies, and other systems with a high degree of structural complexity.  To do so, we combine the physical insights underlying many standard condensed matter order parameters with the mathematical insights provided by the computer science field of ``shape matching.''  We show that virtually all of the standard structural characterization schemes from the more general condensed matter literature can be broken down, fundamentally, into the problem of quantifying the degree to which structures match.  Structural similarity, in turn, can be quantified by using robust ``shape descriptors'' from the field of shape matching, which can be applied to arbitrarily complex structures.  We decompose several order parameters, correlation functions, and other standard structural characterization schemes into their core elements, such that they can be used with arbitrary shape descriptors to extend their applicability.  Additionally, we introduce new, more abstract structural characterization schemes that can also be used with arbitrary shape descriptors, to solve novel problems that arise in computational studies of self-assembly.  The shape matching methods that we provide will facilitate the creation of new structural metrics that are standardized, improving accuracy and comparability, but are also still flexible enough to be applied to the new classes of complex structures that arise in assembly problems.  

This article is organized as follows.  In section~\ref{sec:overview}, we provide an overview of the shape matching framework and terminology that we will employ and describe how it connects to some standard structural characterization schemes from the condensed matter literature.  In section~\ref{sec:descriptors}, we review some relevant shape descriptors from the shape matching literature that can be applied to assembled systems.  In section~\ref{sec:matching}, we introduce some simple ``similarity metrics'' that can be used together with the shape descriptors from section~\ref{sec:descriptors} to measure structural similarity.  Finally, in section~\ref{sec:applications}, we introduce general algorithms based on shape descriptors and similarity metrics that can be used to obtain various types of structural information for complex particle systems.  To demonstrate the usage of these algorithms, we apply shape matching to systems in the fields of  nanoscience, computational self-assembly and condensed matter.   Our examples include identifying local and global structures, quantifying structural changes as a function of time or a control variable, constructing correlation functions, mapping structural phase diagrams, and grouping similar structures.  We cover a wide range of systems including ordered phases formed from spherical and point-like particles, a fluid of tetrahedrally-shaped particles with locally ordered motifs~\cite{amir09}, self-assembled systems of tethered nanoparticles with various nanoparticle shapes~\cite{ditethered, iacovella2009b, tnv}, patchy colloidal tetrominoes~\cite{bubba}, a helical ribbon formed from tethered nanorods~\cite{trunghelix}, a model protein~\cite{ubiquitin}, gold nanowires~\cite{Pu2008,iacovella2010}, and small clusters of water molecules~\cite{water}. The examples that we provide are applicable to particle systems in general, provided that the particle positions and, in some cases, orientations, can be detected.  Although not explicitly treated here, other data representations such as images or diffraction data can also be used to obtain structural metrics within the shape matching framework.  To aid in the development and dissemination of shape matching techniques, we provide accompanying software and examples via the web~\cite{smwebsite}.


\section{Shape Matching Overview}
\label{sec:overview}

\begin{figure}
\begin{center}
\includegraphics[width=1\columnwidth]{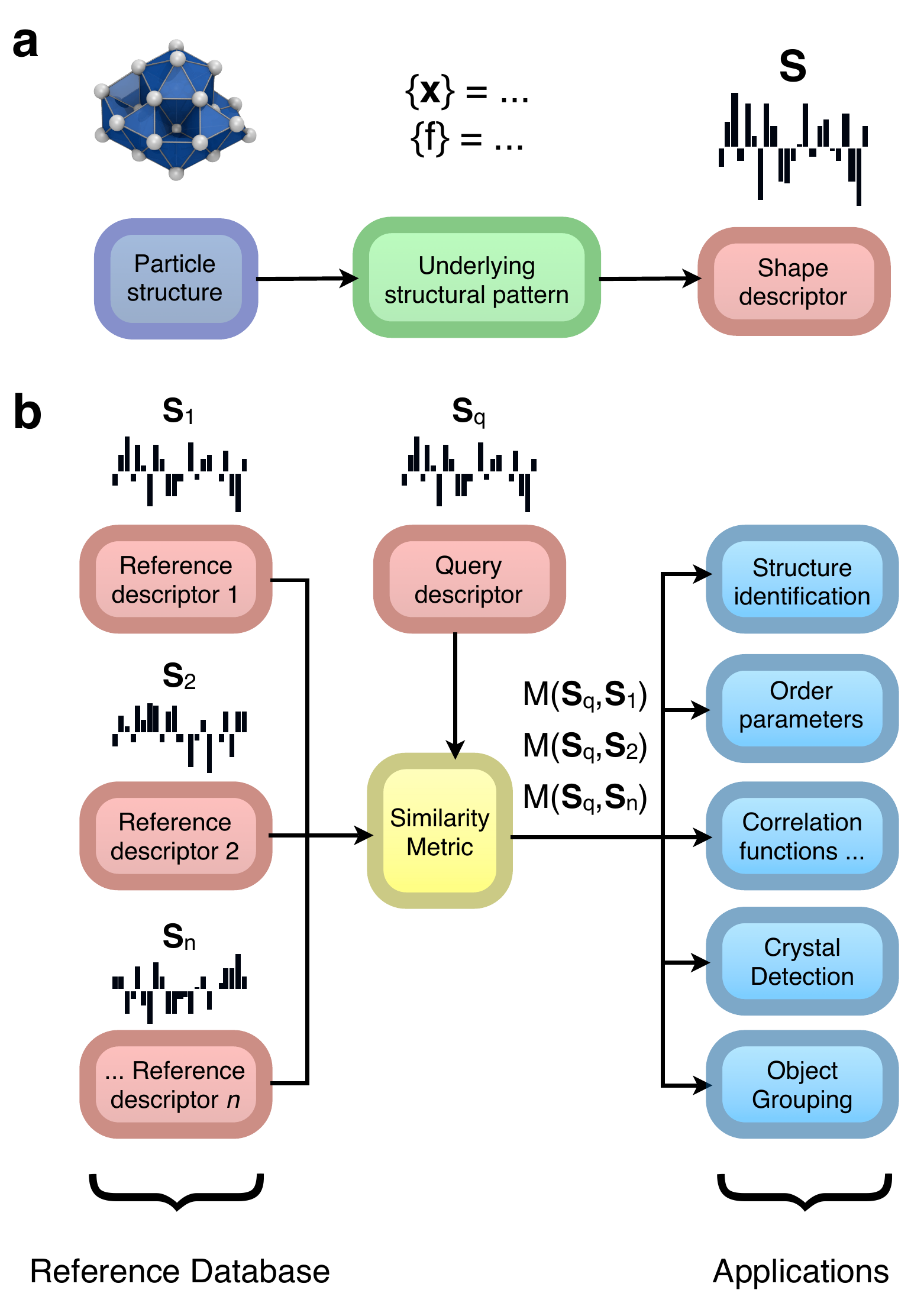}
\caption[Data Flow Diagram for Shape Matching]{Data flow diagram for shape matching.  (\textit{a}) A representative pattern is extracted for a given particle structure and then indexed into a structural fingerprint known as a shape descriptor, $\textbf{S}$.  The depicted cluster is an energy-minimized quantum Lennard-Jones cluster~\cite{ccd, calvo01}. (\textit{b}) Shape descriptors are then compared to obtain similarity information $M$, which can be applied within the context of various structural characterization schemes.}
\label{fig:sldfd}
\end{center}
\end{figure}

The problem of quantifying how well structures match (see Fig.~\ref{fig:sldfd}) has been generalized within the context of the computer science field of ``shape matching''~\cite{veltcamp}.  Familiar shape matching applications include matching fingerprints, signatures~\cite{veltcamp}, faces~\cite{lu2004}, and iris patterns~\cite{ma2004efficient}.  Shape matching schemes have already been applied to systems of particles, particularly in the realm of fast database searches for proteins and macromolecules~\cite{lesk86, ssm, proteinpmpnas, yeh, venkatraman, mak, grandison}.   In some specific cases, shape matching schemes have been explicitly applied to local structure identification in problems in condensed matter and nanoscale self-assembly~\cite{iac07,gyroid,keys07}.   Recently, we  generalized the concept of applying shape matching methods to assembled systems~\cite{keys10-ARCMP}, and demonstrated how a particular class of  harmonic shape descriptors can be applied to a wide variety of self-assembly problems~\cite{keys10-HO}.  In the present article, we provide a thorough survey of both the different types of shape descriptors and structural characterization schemes that can be applied to assembled systems.


The basic idea of shape matching is to ``index'' structures into mathematical fingerprints known as ``shape descriptors,'' $\textbf{S}$, and then compare them using a similarity metric $M(\textbf{S}_i, \textbf{S}_j)$ to obtain both a quantitative and qualitative measure of similarity between the structures.  For mathematical simplicity, we constrain our shape descriptors here to be vectors containing an arbitrary number of components, and our similarity metrics to be scalars that indicate the degree of correspondence between pairs of shape descriptors vectors.  Matching can then be performed using straightforward vector operations, based on, e.g., the degree of alignment of or distance between shape vectors.  Matching information is used to create order parameters and correlation functions, or to identify structures by comparing ``query'' structures to ``reference'' structures.   Since we can choose virtually any structure as a reference, this scheme facilitates the creation of highly specific structural metrics.  The workflow for an application within the shape matching framework is shown in Fig.~\ref{fig:sldfd}.

To apply these ideas to particle systems, we begin by asserting that most standard structural metrics include an implicit concept of ``matching.''  That is, an order parameter or correlation function typically tells us the degree to which a structure of interest matches another (often ideal) structure.  Most standard structural characterization schemes implicitly fit within the shape matching framework, and can be decomposed into query structures, reference structures, shape descriptors, and similarity metrics.  

For example, consider the well-known order parameter $\bar{P}_2$ which detects nematic (aligned) liquid crystalline ordering:
\begin{equation}
\label{eq:nematic}
\bar{P}_2 = \left< P_2(\cos \theta) \right> = \left< \frac{3 \cos^2 \theta-1}{2} \right>.
\end{equation}
The function $P_{2}$ is the second Legendre polynomial~\cite{mathbook} and $\theta$ is the angle between the axis of the molecule and the ``local director'' $\textbf{d}$ that indicates the preferred direction of the overall sample~\cite{larson}.  As the direction deviates from the preferred direction, $P_{2}$ decreases proportionally.  The global nematic order parameter $\bar{P}_2$ is obtained by computing the ensemble average of $P_{2}$, denoted by angle brackets.  In this scheme, the query structure is the system of particles and the reference structure is an ideal nematic liquid crystal with director $\textbf{d}$.  The shape descriptor is given by the collection of angles between the molecular axes and $\textbf{d}$ and the similarity metric is given by the Legendre polynomial $P_{2}$.  The order parameter $\bar{P}_2$ gives an optimal value of $1$ when the structure matches a perfectly aligned liquid crystal with director $\textbf{d}$, and tends toward zero the more the structure deviates from this ideal case.  

As another classical example, consider the hexatic correlation function for 2d systems of spherical particles, or disks ~\cite{halperin78, nelsonc6}:
\begin{equation}
g_{6} (r) = \left<  \frac{  \sum_{i \neq j}  \psi_{6}(i)\psi^*_{6}(j) \delta( r- |\textbf{r}_{i} - \textbf{r}_{j}| )}{ \sum_{i \neq j} \delta( r- |\textbf{r}_{i} - \textbf{r}_{j}| ) } \right> .
\end{equation}
The quantity $\psi_{6}$ is a  ``bond order'' parameter, defined as~\cite{mermin1968, halperin78}:
\begin{equation}
\psi_{6}(i) = \frac{1}{n} \sum_{j}^{n} \exp(i6\theta_{j}).
\end{equation}
Here, $n$ is the number of atoms in the first neighbor shell of an atom $i$, and $\theta$ is the direction of neighboring atom $j$.  The value of $\psi_{6}(i)\psi^{*}_{6}(j)$ approaches $\sqrt{2}$ when the particles $i$ and $j$ are both in hexagonal local environments with the same spatial orientation, and varies towards $0$ otherwise.  Thus, like the standard radial distribution function $g(r)$,  $g_6(r)$ measures the degree of spatial ordering; however, whereas $g(r)$ is sensitive to translational ordering generally, $g_6(r)$ is specifically sensitive to aligned hexagonal ordering.   In this scheme, the query and reference structures are the pairs of neighbor shell clusters for atoms $i$ and $j$, which are a distance $r$ apart, the shape descriptors are the local values of $\psi_{6}$, and the similarity metric is the `dot' operator, which measures the coherence between two $\psi_6$ descriptors.  Thus, $g_{6}(r)$ measures how closely the local environment of a particle at the origin matches with that of a particle a distance $r$ away in terms of both hexagonal shape and spatial orientation. 

Although $\bar{P}_2$ and $g_6(r)$ are specific schemes, the physical insights underlying them are general.  Recasting standard schemes within the shape matching framework allows us to obtain the same types of information, but with different shape descriptors $\textbf{S}$ and similarity metrics $M$ that are better suited for the unique and complex structures observed in assembled systems.  The latter provides the main substance of this article, but first, we introduce several shape descriptors and similarity metrics in the following sections.

\section{Shape Descriptors}
\label{sec:descriptors}

The shape descriptors that we describe in this section are adapted from the computer science field of shape matching.   We constrain our discussion here to the subset of shape matching methods that we believe are most readily applicable to particle systems in both two and three dimensions.  For a more comprehensive review of shape matching methods, see, for example, references~\cite{iyerreview, tangelder, zhangreview}.  

The first step towards creating an order parameter within the shape matching framework is to index the shapes representing the structure of interest into one or more shape descriptors $\textbf{S}$.  For simplicity, we consider in our framework shape descriptors to store structural information in a vector, which may contain real or complex components.  However, shape descriptors may take other forms. 

In addition to containing structural information, shape descriptors may possess other desirable properties and contain additional data, which may determine which descriptor is optimal for a particular application.  One important property of shape descriptors is ``invariance,'' defined as the ability for the descriptor to remain unchanged under certain mathematical transformations, such as scaling, translations, or rotations.  In the context of particle systems, rotation invariance is a highly desirable property, since many applications involve comparing structures in a way that is independent of their spatial orientation.  For descriptors without rotation-invariance, alignment or ``registration~\cite{icp, pca}'' algorithms must be employed prior to matching to remove orientational dependence.  Since particle systems often exhibit thermal noise, another desirable property of shape descriptors is robustness under small perturbations.  However, this property must be balanced with the property of sensitivity, so that descriptors are still capable of detecting subtle structural differences.  Another important consideration is the amount of computational time required to compute and compare the descriptors, which may vary drastically for different schemes.  Often, there is a direct tradeoff between computational cost and accuracy and attention to detail.  
  
In the following sections, we provide a brief overview of some shape descriptors with different combinations of these properties that are well suited for self-assembled systems of particles. These descriptors are not representative of the full realm of possibilities, but rather are meant to serve as demonstrative examples.  It is important to note that in principle, there is no limit on how the shape descriptor is calculated.  Here, we constrain our analysis to descriptors that can be described mathematically as a vector, since this simplifies the process of writing general similarity metrics in section~\ref{sec:matching}.  However, in general, not all descriptors can be represented in this way, and thus require different similarity metrics. 


\begin{figure*}
\includegraphics[width=0.75\textwidth]{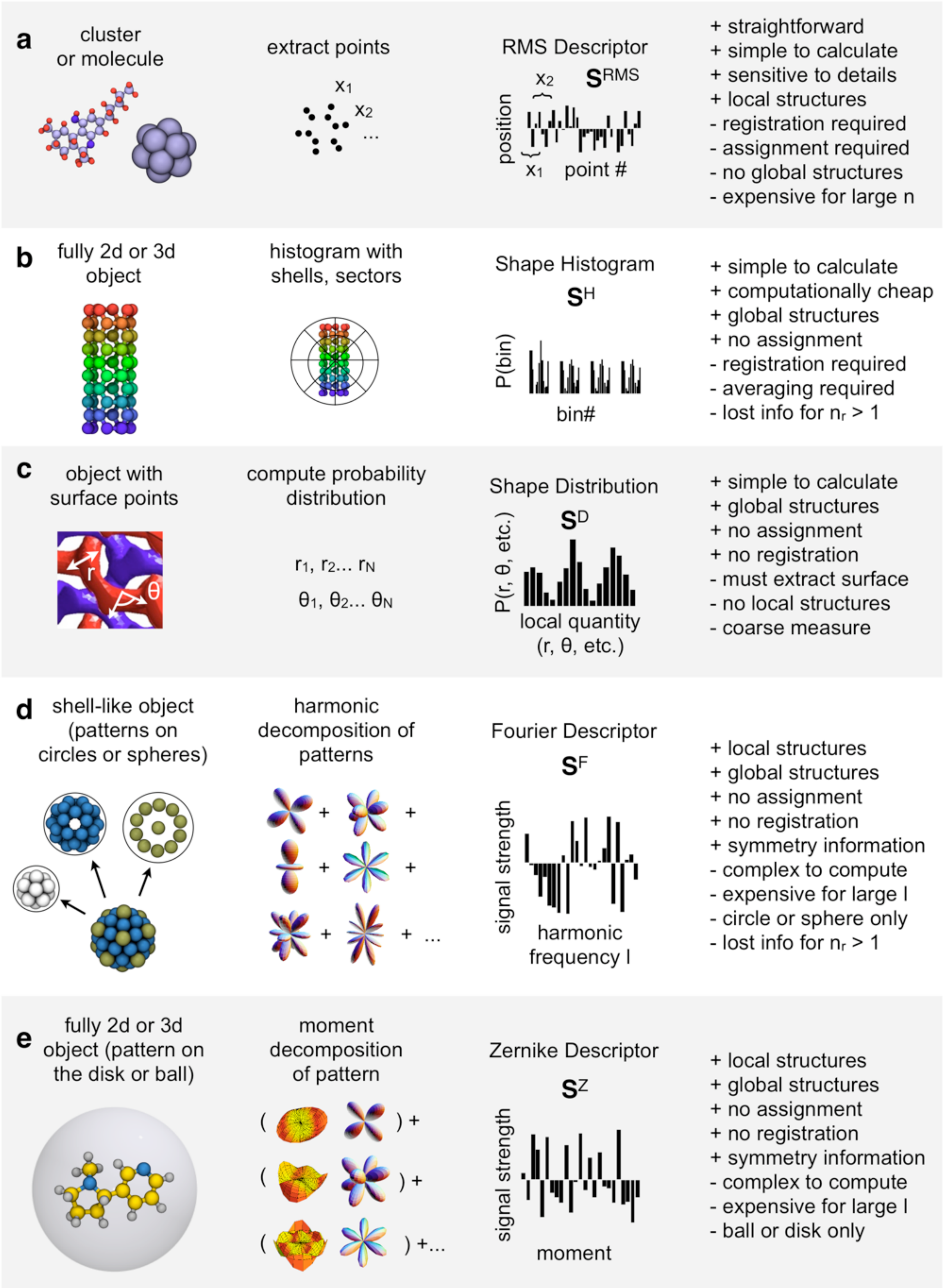}
\caption[Depiction of Five Different Shape Descriptors]{Depiction of five different shape descriptors. (\textit{a}) The RMS descriptor~\cite{nrpointmatching, icp}.   Descriptor components are given trivially by particle positions or density map.  (\textit{b}) The shape histogram descriptor~\cite{ankerst}.  The structure is indexed into a histogram consisting of $n_{r}$ shells and $n_{\theta}$ sectors.  (\textit{c}) The D2 shape distribution descriptor~\cite{osada}.  The probability distribution is computed for various local measurements, such as the distance or angle between surface points.  (\textit{d}) The Fourier descriptor~\cite{fourier, ylm}.  A pattern along the perimeter of the circle or on the surface of a sphere is decomposed into a harmonic representation. (\textit{e}) The Zernike descriptor~\cite{zernike2d, zernike3d}.  A pattern on the unit disk or unit ball is decomposed into a harmonic representation.}
\label{fig:sldescriptors}
\end{figure*}

\subsection{Data Representations}
Particle systems are typically represented as either a set of points (point cloud data) or solid objects (volumetric data).  Both types of data can be represented by a set of position vectors $\{X\} = \{\textbf{x}_1, \textbf{x}_2, ... \textbf{x}_n\}$ and weights $\{f\} = \{f_1, f_2, ... f_n\}$.  Point cloud data $\{X\}$ typically represents particle positions, in which case the weights $f_i$ are all 1.  For volumetric data, the position vectors $\textbf{x}_i$ represent the location of voxels (n-dimensional pixels) with intensities given by $f_i$.   There is no formal rule regarding how to best represent input data for a given system.  In general, point cloud data is optimal when particle shapes are not important, such as is the case with point particles.   Volumetric data is optimal when particle size, shape, or orientation are important, such as with systems of rods or polyhedra, or when the system is coarse-grained in space, such as with phase-separated structures.  Image processing algorithms ~\cite{varadan2003, crocker1996,ditethered} can often be employed to change between the two data representations.

\subsection{Point-Matching or RMS Descriptor}
\label{ssec:RMS}
For relatively simple structures, such as small clusters or macromolecules, we can use the particle positions themselves as a shape descriptor (Fig.~\ref{fig:sldescriptors}a).  Matching for this simple scheme is often based on the root-mean-square (RMS) difference between points, and thus the scheme itself is often referred to as ``RMS matching.''  Mathematically, the point matching descriptor $\textbf{S}^\mathrm{RMS}$ is defined trivially by the pointset $\{X\}$:
\begin{equation}
\textbf{S}^{\mathrm{RMS}} = \{\textbf{x}_1, \textbf{x}_2, ... \textbf{x}_n\}.
\end{equation}
Here, each $\textbf{x}_i$ is a $d$-dimensional vector representing the position of the $i$th point in $\{X\}$.  The $\textbf{S}^\mathrm{RMS}$ shape descriptor ~\cite{lesk86, nrpointmatching, icp} is a vector with $n \times d$ components.  Typically the centroid is subtracted off and the vectors in $\{X\}$ are normalized, e.g. by dividing by the average distance between points.  Point matching schemes were applied in early attempts at shape-based database searches for macromolecules~\cite{lesk86}, and increasingly powerful variations of these schemes have since been implemented~\cite{proteinpmpnas}.  Although point matching schemes have the advantage of being conceptually simple, there are many subtle drawbacks associated with them.  First, point matching requires an assignment step to determine the optimal correspondence between points in compared structures.  The coordinates in the shape descriptors are then re-ordered accordingly.  As a coarse approximation, points can be assigned based on the minimum distance or the maximum alignment between individual coordinates.  For example, a point $i$ on the query structure, can be assigned to the point $j$ on the reference structure that maximizes the fitness $w_{i,j}$, defined as, e.g., $w_{i,j} = f_i f_j |\textbf{x}_i \cdot \textbf{x}_j|$.  This scheme has the disadvantage that it is possible to assign multiple points on the query structure to a single point on the reference structure.  A more robust method involves creating a ``fitness matrix'' that records the degree of correspondence between all pairs of points:
\begin{equation}
F =  \left( \begin{array}{ccc}
w_{1,1} & w_{1,2} & \dots w_{1, n_{r}} \\
w_{2,1} & w_{2,2} & \dots w_{2, n_{r}} \\
\vdots & \vdots & \vdots \\
w_{n_{q}, 1} & w_{n_{q}, 2} & w_{n_{q}, n_{r}} \end{array} \right) .
\end{equation}
The variables $n_q$ and $n_r$ represent the number of points in the query and reference structures, respectively.  We can then use a numerical technique, such as the Hungarian method~\cite{hungarian}, to efficiently determine the optimal assignment matrix that maximizes the overall fitness of the match.  An additional subtlety arises when $n_q \neq n_r$.  In this case, outliers can be excluded to obtain a ``partial match'' between structures.  This is accomplished by sequentially removing points with the lowest total fitness $w_i$, defined as $w_i = \sum_{j=1}^n w_{i,j}$.  The number of points excluded depends on the desired application.  For partial matching, we might exclude $|n_{q} - n_{r}|$ points from whichever structure contains the fewest points.  For excluding outliers, we might exclude all points with $w_i$ below a certain threshold.

In addition to requiring assignment, the RMS descriptor also has the drawbacks that it is sensitive to scale, position, and orientation, and structures must first be normalized and registered unless the orientations are known beforehand or the intended application utilizes rotation-dependent matching.  Depending on the application, objects may be registered based on rigid alignment, or other constraints~\cite{nrpointmatching}.  Rigid registration can be achieved using either the iterative closest point (ICP) method~\cite{icp}, which involves minimizing the distance between points on compared objects by iterative rotations and translations, or the principle components analysis (PCA) method~\cite{pca}, which aligns objects with common principle axes.  The ICP method has the disadvantage that it is non-trivial to implement, computationally expensive for structures with many points, and must be performed for all pairs of compared shapes.  Moreover, it is prone to error if applied naively; the ICP method converges to a local minimum, so many initial orientations need be attempted to ensure convergence to a global minimum.  The PCA method is only applicable to objects with distinct principle axes and thus fails for spherical objects. Despite the simplicity of the point-matching shape descriptor, implementation of the RMS method can often be non-trivial.  Since both assignment and registration are computationally expensive (i.e. they scale poorly with $n$) point matching descriptors should be avoided unless (1) $n$ is small, (2) matching is required for only a few structures, or (3) registration is not required.

\subsection{Shape Histogram Descriptor}
\label{ssec:shapehist}
Another shape descriptor that is conceptually simple and has been applied to molecular database searches is known as the ``shape histogram''~\cite{ankerst} (Fig.~\ref{fig:sldescriptors}b).  This descriptor is based on a density map of the structure on a polar or spherical grid.  The shape histogram is constructed in 2d by first generating $n_\theta$ equiangular gridlines on the unit circle:
\begin{equation} 
\theta_i = 2\pi i / n_{\theta} \quad \quad i = 0, \dots, n_{\theta}-1.
\end{equation}
The value of $n_\theta$ is chosen so as to capture important structural features while balancing computational efficiency.  Structures with radial dependence can be divided into $n_r$ concentric shells.  A given component in the 2d shape histogram descriptor is then given by:
\begin{equation}
S^{\textrm{H2}}_{jn_\theta+k} = \sum_{i=1}^n { f_i 
\delta \left( \bigg\lfloor \frac{n_r |\textbf{x}_i|}{r_{max}} \bigg\rfloor  - j \right) } \delta \left( \bigg\lfloor \frac{n_\theta \theta(\textbf{x}_i)}{2\pi} \bigg\rfloor - k \right).
\end{equation}
The $\textbf{S}^\mathrm{ H2}$ descriptor then contains $n_\theta n_r$ real components, one for each bin in the histogram:
\begin{equation}
\textbf{S}^{\textrm{H2}} = \left< S^{\textrm{H2}}_1, S^{\textrm{H2}}_2, \dots S^{\textrm{H2}}_{n_\theta n_r} \right>.
\end{equation}

The 3d version of the shape histogram is constructed in a similar way, except that in this case, there are many different ways to construct the grid.  An equiangular grid with and $n_\theta$ azimuthal bins and $\frac{1}{2}n_\phi$ polar bins is given by:
\begin{equation}
\theta_i = \pi i / n_\theta, \quad  \phi_j = 2\pi k  / n_\phi; \\
\end{equation}
\begin{equation}
\quad i=0, 1, \dots n_\theta-1, \quad j=0, 1, \dots, n_\phi-1. \nonumber
\end{equation}
The total number of cells defined by the gridlines is $\frac{1}{2}n_\phi^2$.   The 3d equiangular grid introduces artifacts near the poles of the sphere where the cells are small compared to the equator.  Such artifacts are inherent to 3d grids on the sphere; there is no way to create an evenly-spaced grid on the sphere with equivalent cells.  However, there are several alternatives to the equiangular grid, such as the rectilinear grid, icosahedral grid, etc., that give more evenly-sized cells~\cite{spheregrid}.  A given component in the 3d shape histogram descriptor for an equiangular grid is given by:
\begin{widetext}
\begin{equation}
S^{\textrm{H3}}_{ \substack{ jn_\theta n_\phi \\ +  kn_\phi \\ + l} } = \sum_{i=1}^n  f_i 
\delta \left( \bigg\lfloor \frac{n_r |\textbf{x}_i|}{r_{max}} \bigg\rfloor - j \right) 
\delta \left( \bigg\lfloor \frac{n_\theta \theta(\textbf{x}_i)}{\pi} \bigg\rfloor - k  \right)
\delta \left( \bigg\lfloor \frac{n_\phi \phi(\textbf{x}_i)}{2\pi} \bigg\rfloor - l  \right).
\end{equation}
\end{widetext}
As for $\textbf{S}^{\mathrm{H2}}$, shapes with $r$-dependence are indexed by computing separate angular histograms for each radial shell.  The $\textbf{S}^\mathrm{H3}$ descriptor contains $\frac{1}{2}n_\phi^2 n_r$ real components, one for each bin in the histogram:
\begin{equation}
\textbf{S}^{\textrm{H3}} = \left< S^{\textrm{H3}}_1, S^{\textrm{H3}}_2, \dots S^{\textrm{H3}}_{n_r n_\phi^2 / 2} \right>.
\end{equation}
The shape histogram has several advantages over the point matching method.  First, no assignment step is required, since the histogram does not retain information about the ordering of the input points.  Additionally, the grid resolution can be adjusted by modifying $n_\theta$ and/or $n_\phi$ to provide a desired degree of spatial coarse-graining.  The shape histogram has the disadvantage that, like the point matching method, it requires registration to match shapes that are not aligned spatially, unless only radial bins are used (i.e., $n_\theta = n_\phi = 1$).   Shape histograms with only radial bins are typically only applicable for obtaining coarse measures of similarity, since shape histograms lose much of their discerning capabilities without an angular component.  If $n$ is large, the cost of registration can be significantly reduced by aligning the histograms themselves rather than the raw data.  Shape histograms are best suited for describing structures that can be broken down into concentric circles or spheres.  Examples include nanoparticle clusters, proteins and macromolecules.  Shape histograms are also well suited for indexing global structures with orientational ordering such as crystals or quasicrystals, wherein the bond or neighbor directions of particles create a global pattern on the circle or sphere, as described in section~\ref{ssec:global}.  

\subsection{Shape Distributions}
\label{ssec:shapedist}
For many applications, registration is costly and rotation-invariant descriptors are optimal.  A simple yet powerful method for creating rotation-invariant descriptors is given by the ``shape distributions'' scheme~\cite{osada} (Fig.~\ref{fig:sldescriptors}c).   This scheme involves creating distribution functions for simple invariant local metrics.  The shape distribution ``D2'' is defined as the probability distribution of observing two surface points $i$ and $j$ a distance $r$ apart.  A given component in the D2 descriptor is given by:
\begin{equation}
S^{\textrm{D2}}_k = \sum_{i \neq j} f_i f_j \delta \left[ \bigg\lfloor \frac{ n_r \left(\left| \textbf{x}_i - \textbf{x}_j \right| - r_{min}\right)}{r_{max}-r_{min}} \bigg\rfloor - k \right].
\end{equation}
The D2 descriptor is the collection of $n_r$ radial components:
\begin{equation}
\textbf{S}^{\textrm{D2}} = \left<  S^{\textrm{D2}}_1, S^{\textrm{D2}}_2, \dots S^{\textrm{D2}}_{n_r} \right>.
\end{equation}
Notice that this function is similar to the standard radial distribution function $g(r)$, except that there is no ideal gas normalization and the function is typically computed only for points on the surface of the object.

A similar distribution ``A3'' is defined by the probability of observing an angle $\theta$ between three surface points:  
\begin{equation}
S^{\textrm{A3}}_l = \sum_{i \neq j \neq k} f_i f_j f_k \delta \left( \bigg\lfloor \dfrac{ n_\theta \textbf{x}_j \cdot \textbf{x}_k} { \pi \sqrt{( |\textbf{x}_j - \textbf{x}_i | | \textbf{x}_j - \textbf{x}_i |)} } \bigg\rfloor - l \right).
\end{equation}
The A3 descriptor is the collection of $n_{\theta}$ components:
\begin{equation}
\textbf{S}^{\textrm{A3}} = \left<  S^{\textrm{A3}}_1, S^{\textrm{A3}}_2, \dots S^{\textrm{A3}}_{n_\theta} \right>.
\end{equation}
Notice that this function is similar to the angular distribution function $a(\theta)$.  

Similar distributions can be contrived for sets of four, five, etc., points; however, D2 and A3 were shown to have the best discerning capabilities for the structures tested in reference~\cite{osada}.  Shape distributions are best applied to structures with clearly defined, distinguishable surfaces, such as phase-separated structures formed by block copolymers~\cite{dotera, reister} or tethered nanoparticles~\cite{glotzer2005, zhang2003, tns, tnv}.  Like $g(r)$ and $a(\theta)$, shape distributions are too coarse to distinguish between similar shapes, such as small polyhedral clusters.

\subsection{Fourier Descriptors}
\label{ssec:fourier}

For shapes with more subtle differences, such as  localized nanoparticle clusters, macromolecules, or global crystal structures, we can apply a more complex but more powerful technique for creating rotation invariants based on computing the harmonic transform of the shape histogram.  By disregarding phase information from the harmonic transform, we obtain descriptors that are invariant under rotations. The formulae for the harmonic transform depend on the underlying basis.  Invariants can be obtained for shapes on the unit circle~\cite{fourier} ($\theta$-dependence), sphere~\cite{ylm, ylm2} ($\theta,\phi$-dependence), disk~\cite{zernike2d} ($r, \theta$-dependence) or ball~\cite{zernike3d} ($r,\theta,\phi$-dependence).  On the unit circle or sphere, the harmonic descriptors are known as Fourier descriptors (Fig.~\ref{fig:sldescriptors}d).  On the unit disk or ball, the descriptors are known as Zernike descriptors (Fig.~\ref{fig:sldescriptors}e), which we discuss in the following section.  Additional details regarding the properties and  implementation of these descriptors for molecular systems are provided in a separate reference~\cite{keys10-HO}.  

The Fourier descriptors are based on the Fourier transform, which involves decomposing a function into a sum of harmonic components.  The Fourier coefficients for a 2d pattern are obtained by computing the discrete Fourier transform for each ``shell'' $s$ of the 2d shape histogram, $\textbf{S}^{\textrm{H2}}$, defined in section~\ref{ssec:shapehist}:  
\begin{equation}
\psi_{\ell,s} = \dfrac{\sum_{j=0}^{n_{\theta}-1} S^{\textrm{H2}}_{sn_{\theta}+j} \exp{\left(-i \ell \frac{ 2\pi j}{n_{\theta}} \right)}}{\sum_{j=1}^{n_\theta} S^{\textrm{H2}}_{sn_{\theta}+j}}.
\end{equation}
Here, $n_\theta$ is the number of sectors in each shell $s$ in the shape histogram.  By considering each shell independently, we reduce a 2d problem (a function of $r$ and $\theta$) to $n_r$ 1d problems (functions of $\theta$ only).  The coefficients $\psi_\ell$ are complex numbers.  

Although the Fourier coefficients in their complex number form are not rotation-invariant (which may be beneficial for some applications), they can be converted to an invariant form by computing the magnitude of each coefficient.  The invariant coefficients for a pattern on the circle are given by:
\begin{equation}
|\psi_\ell| = \psi_\ell \psi_\ell^* = \left[\Re(\psi_\ell)^2 + \Im(\psi_\ell)^2\right]^{1/2}.
\end{equation}
The Fourier invariants are positive real numbers.  To create a Fourier descriptor for a given shell $s$, we take a collection of desirable coefficients: 
\begin{equation}
\textbf{S}^{\mathrm{F2}}_s = \left< |\psi_{\ell_{min}, s}|, |\psi_{\ell_{min}+1, s}|, \cdots |\psi_{\ell_{max}, s}| \right> .
\end{equation}
In this equation, we have used invariant coefficients; however, rotation-dependent coefficients are useful for many applications~\cite{snr83, tenwolde96, auer04, keys07}.  The coefficients are sensitive to patterns with angular frequencies that match the parameter $\ell$.  For example, $\psi_4$ is large for 4-fold patterns, $\psi_6$ is large for 6-fold patterns, etc.  Specific coefficients can be chosen to describe structures with particular angular frequencies.  In general, an arbitrary pattern can always be described by a sufficiently large range of $\ell$.  For the problems that we consider, we typically take $\ell$ in the range $\ell_{min} \sim 2$, $\ell_{max} \sim 10$.  The overall Fourier descriptor is given by concatenating the descriptors for each shell into a vector:
\begin{equation}
\textbf{S}^{\mathrm{F2}} = \left< \textbf{S}^{\mathrm{F2}}_1, \textbf{S}^{\mathrm{F2}}_2, \dots \textbf{S}^{\mathrm{F2}}_{n_r} \right> .
\end{equation}

An analogous scheme can be used for 3d objects, where shells in the shape histogram have $[\theta, \phi]$ dependence.  The Fourier coefficients are obtained by computing the discrete spherical harmonics transform for each ``shell'' $s$ of the 3d shape histogram, $\textbf{S}^{\textrm{H3}}$:
\begin{widetext}
\begin{equation}
\textbf{q}_{\ell, s}  = 
\dfrac{\sum_{j=0}^{n_\theta-1} \sum_{k=0}^{n_\phi-1} S^{\textrm{H3}}_{\substack{s n_\theta n_\phi\\ + j n_\phi\\ + k} } N_\ell^m P_\ell^m \left[ \cos \left( \frac{\pi j}{n_\theta} \right) \right] 
\exp{ \left (-i m \frac{2 \pi k}{n_\phi} \right)} }
{ \sum_{j=1}^{n_\theta} \sum_{k=1}^{n_\phi}  S^{\textrm{H3}}_{\substack{s n_\theta n_\phi\\ + j n_\phi\\ + k} } }.
\end{equation}
\end{widetext}
Here, $N_\ell^m$ is a normalization factor $\sqrt{{(2\ell+1) }{(\ell-m)! / (\ell+m)!}} $, and $P_\ell^m$ is a Legendre polynomial~\cite{mathbook}.  The variable $m$ is an integer $m \in [-\ell, \ell]$.  Therefore, unlike the circular coefficients $\psi_\ell$, which are complex numbers, the spherical coefficients $\textbf{q}_\ell$ are vectors with $2\ell+1$ complex components.  Rotation-invariant versions of the coefficients can be obtained by computing the vector magnitude:
\begin{equation}
|\textbf{q}_\ell| =  \left( \frac{4\pi}{2\ell+1}\sum_{m=-\ell}^\ell |q_\ell^m|^2 \right) ^{1/2}.
\end{equation}
Like the Fourier invariants on the circle $|\psi_\ell|$, the Fourier invariants on the sphere $|\textbf{q}_\ell|$ are positive real numbers.  To create a Fourier descriptor for a given shell $s$, we take a collection of desirable coefficients: 
\begin{equation}
\textbf{S}^{\mathrm{F3}}_s = \left< |\textbf{q}_{\ell_{min}, s}|, |\textbf{q}_{\ell_{min}+1, s}|, \cdots |\textbf{q}_{\ell_{max}, s}| \right> .
\end{equation}
Again, we have chosen invariant coefficients, but rotation dependent coefficients may also be used.  The overall Fourier descriptor on the unit sphere is given by:
\begin{equation}
\textbf{S}^{\mathrm{F3}} = \left< \textbf{S}^{\mathrm{F3}}_1, \textbf{S}^{\mathrm{F3}}_2, \dots \textbf{S}^{\mathrm{F3}}_{n_r} \right> .
\end{equation}
Again, different combinations of coefficients can be used to create shape descriptors with different levels of robustness and sensitivity to particular symmetries. 

By using harmonic descriptors we gain many of the same advantages of the shape histogram, but without the need to register the objects or histograms.  Like the shape histogram, harmonic descriptors are well suited for describing a wide variety of shapes including nanoparticle clusters, proteins and macromolecules, crystals composed of arbitrarily shaped particles and, in some cases phase separated structures.  Harmonic descriptors exhibit an inherent data smoothing mechanism; thus they are typically better-suited for describing small polygonal or polyhedral clusters than the shape histogram, which is prone to error without sufficient averaging.  These properties, along with the unique ability to yield symmetry-specific information, have already been successfully applied to constructing orientational order parameters for small clusters of point particles and simple crystals in the context of bond order parameters~\cite{halperin78, nelsonc6, snr83, mermin1968}.  While the bond order parameters scheme focuses primarily on the numerical values of specific coefficients (often $\psi_6$ and $q_6$), the shape matching approach more closely resembles a signal processing application where we utilize a broad spectrum of Fourier coefficients.  Additionally, while the bond order parameters were defined for point clusters that form patterns on the circle or sphere, the descriptors introduced here can be applied to volumetric objects and objects with $r$-dependence.  Notice that in the limit of infinitesimal angular bin size and a single radial bin ($n_\theta, n_\phi \rightarrow \infty$, $n_r = 1$), our definitions of $\psi_\ell$ and $\textbf{q}_\ell$ become nearly equivalent to the bond order parameters, only differing by a sign in the complex exponential.  This only makes the direct mathematical connection between harmonic descriptors and the Fourier transform more explicit; the change is otherwise inconsequential.  We explore the properties of Fourier descriptors in more detail in reference~\cite{keys10-HO}.


\subsection{Zernike Descriptors}

The Fourier descriptors introduced in the previous section have pseudo $r$-dependence.  That is, radial information is incorporated by decomposing the structure into concentric shells and then computing independent descriptors for each shell.  This is problematic for structures with a small number of sample points, such as small clusters, because random perturbations can move points between nearby shells.  A second, more subtle drawback occurs when attempting to distinguish between structures for which the shapes of the shells are similar, but the relative orientation of the shells within the structure are different\cite{ylm}.  Since the descriptors are computed for each shell independently, rotation-invariant descriptors are insensitive to relative orientations within the structure~\cite{ylm}.  As a result, in many cases, it is preferable to compute harmonic descriptors with full $r$ dependence, known as ``Zernike descriptors~\cite{zernike2d, zernike3d}'' (Fig.~\ref{fig:sldescriptors}e).  The coefficients of the Zernike expansion, known as ``Zernike moments,'' are computed by adding a Zernike radial polynomial to the Fourier coefficients: 
\begin{equation}
R^m_n(r) = \! \sum_{k=0}^{(n-m)/2} \!\!\! \frac{(-1)^k\,(n-k)!}{k!\,((n+m)/2-k)!\,((n-m)/2-k)!} \;r^{n-2\,k}.
\end{equation}
Here, $r$ is the radial distance from the origin $r \in [0, 1]$, and $m$ and $n$ are integers, $n \geq m > 0$.  The Zernike moments on the 2d unit disk are given by:
\begin{equation}
a_{n \ell} = \dfrac{ (n+1)
\sum_{j=0}^{n_{r}-1} \sum_{k=0}^{n_{\theta}-1} S^{\textrm{H2}}_{jn_{\theta}+k} R_{n \ell}\left( \frac{j}{n_r} \right) \exp{\left(-i \ell \frac{ 2\pi k}{n_{\theta}} \right)} }
{\pi \sum_{j=1}^{n_r} \sum_{k=1}^{n_\theta} S^{\textrm{H2}}_{j n_{\theta}+k}} .
\end{equation}
The moments are subject to the constraint that $\ell \leq n$ and $(n - \ell )$ is even.  Each moment is a complex number.  The rotation invariant Zernike moments on the unit disk are given by:
\begin{equation}
|a_{n\ell}| = a_{n\ell} a_{n\ell}^*.
\end{equation}
The 2d Zernike invariants are positive real numbers.  A Zernike descriptor can be created by concatenating the desired Zernike moments into a vector, for example:
\begin{equation}
\textbf{S}^{\mathrm{Z2}} = \left< |a_{11}|, |a_{20}|, |a_{22}|, \dots |a_{\ell_{max}, \ell_{max}}|   \right> .
\end{equation}
Again, we have chosen invariant coefficients, but rotation dependent moments may also be used.  The Zernike moments on the 3d ball are given by:
\begin{widetext}
\begin{equation}
\textbf{z}_{n \ell} =
\dfrac{3(n+1) \sum_{j=0}^{n_r-1} \sum_{k=0}^{n_\theta-1} \sum_{l=0}^{n_\phi-1} S^{\textrm{H3}}_{ \substack{j n_{\theta} n_\phi\\ + n_\phi k\\ + l}} 
R_{n \ell}\left( \frac{j}{n_r} \right) N_\ell^m P_\ell^m \left[ \cos \left( \frac{\pi k}{n_\theta} \right) \right] 
\exp{ \left (-i m \frac{2 \pi l}{n_\phi} \right) }} 
{4\pi \sum_{j=1}^{n_r} \sum_{k=1}^{n_\theta} \sum_{l=1}^{n_\phi}  S^{\textrm{H3}}_{ \substack{j n_{\theta} n_\phi\\ + n_\phi k\\ + l}} }.
\end{equation}
\end{widetext}
Again, we take $\ell \leq n$ and $(n - \ell )$ is even.  Whereas the 2d Zernike moments $a_{n\ell}$ are  complex numbers, the 3d Zernike moments $\textbf{z}_{n\ell}$ are complex vectors with $2\ell + 1$ components.  The invariant Zernike moments on the unit ball are given by:
\begin{equation}
|\textbf{z}_{n\ell}| =  \sqrt { \frac{4\pi}{2\ell+1}\sum_{m=-\ell}^\ell |z_{n\ell}^m|^2 }.
\end{equation}
The 3d Zernike invariants are positive real numbers.  A 3d Zernike descriptor is created by concatenating the desired moments into a vector, for example:
\begin{equation}
\textbf{S}^{\mathrm{Z3}} = \left< |\textbf{z}_{11}|, |\textbf{z}_{20}|, |\textbf{z}_{22}|, \dots |\textbf{z}_{\ell_{max}, \ell_{max}}|   \right> .
\end{equation}
Again, we have chosen invariant moments, but rotation-dependent moments and other combinations of frequencies may also be used depending on the problem.

Zernike descriptors are best applied to shapes that cannot be described by angles alone, such as certain clusters of nanoparticles, macromolecules, or complex crystals.  When computing Zernike moments it is essential that the patterns being compared are normalized consistently on the unit ball or disk.  Typically, normalization is performed by translating the centroid of the structure to the origin and rescaling the coordinates such that every point on the pattern has a radial distance less than 1.  This scheme is sufficient for the majority of patterns that we encounter in assembled systems.

\subsection{Combined Descriptors}
In many cases, we can create new descriptors by taking linear combinations of the descriptors outlined above.  Since the descriptors are represented as vectors, they can be concatenated together to combine their properties.  Descriptors may also be multiplied by a weighting vector to (de)emphasize certain components.  Other simple descriptor operations, such as averaging or taking probability distributions can also be useful, particularly for describing global structures, as outlined in section~\ref{ssec:global} below.

More complex combinations of descriptors can be created for specific applications.  For example, one powerful solution to the problem of ``partial matching'' is given by the ``shape contexts'' method~\cite{belongie}, which combines elements of the point matching descriptor with the shape histogram descriptor.  A separate shape histogram is computed for each point in the structure, where the coordinate system is centered at that point.  The points in the query structure are then assigned to their corresponding reference structure points by optimizing the match between shape histograms.  Outlier points that do not correspond well can be excluded to obtain a partial match.  Another powerful method based on combining descriptors is given by the light-field descriptor~\cite{lightfield}.  This descriptor combines information obtained from taking several 2D images of a 3D structure at several different vantage points (typically the 20 vertexes of a dodecahedron).  Each 2D image is then indexed by an appropriate shape descriptor, and assignment is performed between the collection of images obtained from different structures to optimize correspondence.  In practice, many initial rotations of the reference frame are attempted to find a rotation that optimizes correspondence.  Although the shape contexts and lightfield descriptors are specialized, the method of combining descriptors to optimize properties is applicable to a wide range of problems.

\subsection{Other Possible Descriptors}
The shape descriptors that we have introduced above are not meant to represent a complete set, but rather representative examples.  Any quantitative measure of structure can be used as a shape descriptor provided that it can be indexed into an $n$-dimensional vector (or matrix).  Along this line, there are several metrics defined in the literature that could fit into the shape matching framework and could be used to inspire useful new shape descriptors.  For example, diffraction patterns, radial distribution functions, or orientation tensors (e.g. radius of gyration tensor or nematic order tensor ~\cite{dijkstra2003}) could be indexed into shape descriptors that describe global structures.  Schemes such as the common neighbor analysis scheme proposed in reference~\cite{cna}, could also be readily incorporated into this framework to describe local structures.  Additionally, other structural metrics from the literature that individually may not be independently distinguishing for a wide range of problems, could still yield useful information through linear combination.

\subsection{Extracting Global Patterns for Shape Descriptors}
\label{ssec:global}
With the exception of shape distributions, the descriptors defined in the preceding sections are designed to index local structures such as small clusters of atoms or nanoparticles, macromolecules, or large but finite micro or nanoscale assemblies.  Describing global structures is more difficult, since local shapes must first be extracted from the infinite system and then combined into patterns that reflect the ``global shape'' for indexing.  The manner in which we construct global patterns depends on the structural properties of the system.  In a rough sense, we can group global structures into two different categories: structures with long range orientational ordering (OO), and those without.

\begin{figure*}
\includegraphics[width=0.80\textwidth]{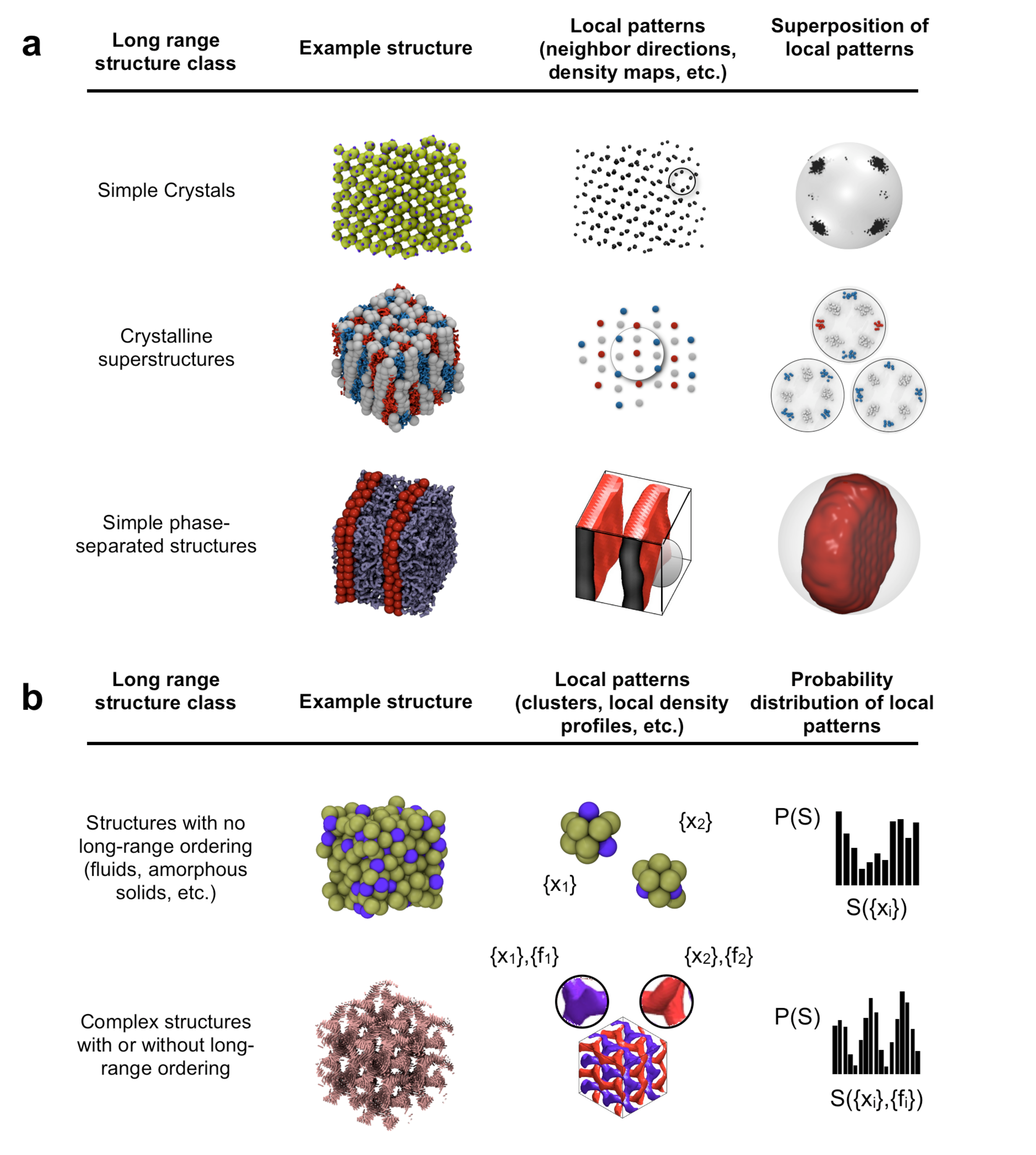}
\caption[Depiction of Strategies for Extracting Global Patterns]{Depiction of strategies for extracting global patterns. (\textit{a}) Global patterns by superposition.   For structures with long range orientational ordering, a global pattern can be extracted by translating all local clusters or density maps to a common origin.  (\textit{b}) For structures with no long range orientational ordering or complex structures with many important directions, a global pattern can be built up from the probability distribution of local patterns.}
\label{fig:slglobal}
\end{figure*}

For structures with long-range OO, such as crystals and quasicrystals~\cite{quasicrystals}, the probability density of neighboring particles is highly correlated for all particles in the system.  Thus, an intuitive global shape is given by the superposition of all neighbor directions for each local structure in the system~\cite{snr83}, sometimes called a ``bond order diagram~\cite{roth2000}.''  This is depicted for the diamond structure~\cite{zhenlidiamond} in Fig.~\ref{fig:slglobal}a, top.  As detailed in the previous sections, this type of pattern is best indexed by the shape histogram, or, for rotation-invariant matching, the Fourier descriptors or Zernike descriptors.  In the case that it is important to distinguish between particle types, independent global descriptors should be created for each type independently, and combined later via concatenation.  An example is given for the tetragonal cylinder structure formed from tethered nanospheres~\cite{ditethered} in Fig.~\ref{fig:slglobal}a, middle.  Global descriptors based on orientational ordering are applicable to crystalline structures in general, including phase-separated systems arranged in crystalline superstructures ~\cite{ditethered, iacovella2009b}.  In this case, the neighbor directions are computed for the centers of the micelles, cylinders, etc. rather than the individual particles.

Non-crystalline globally-ordered phase-separated structures such as layered or network structures can be approached in a similar way.  However, rather than creating a descriptor based on the superposition of local neighbor directions, a global descriptor is built up based on the superposition of local density maps.  An example is given by the lamellar structure formed by tethered nanospheres~\cite{tns} in Fig.~\ref{fig:slglobal}a, bottom.  The resulting patterns can be indexed by shape histograms, Fourier descriptors, Zernike descriptors, etc. in the same way as for crystalline long range order.  To capture ordering on a range of lengthscales, descriptors should be created with a radial component that spans the lengthscales of interest.

For systems with no long range ordering such as liquids, gases and amorphous solids, a different approach must be used.  Since combining neighbor directions or density maps by superposition in non-distinguishing, we instead compute the probability distribution of these local patterns.  This method is depicted for a dense liquid~\cite{KA} in Fig.~\ref{fig:slglobal}b.  Since this requires a separate descriptor for every local structure, registration becomes computationally prohibitive.  Thus, rotation-invariant descriptors, such as Fourier descriptors or Zernike descriptors, are typically optimal.  Computing probability distributions is also useful for complex structures regardless of long range ordering.  For example, for the double gyroid structure shown in Fig.~\ref{fig:slglobal}b, bottom~\cite{gyroid}, the superposition of local density maps may become non-distinguishing for the global sample since there are many different directions, and probability distributions may present a better alternative.  As mentioned in section~\ref{ssec:shapedist}, complex phase-separated structures can often be distinguished by shape distribution descriptors.  However, while these descriptors are simple, they yield only a coarse measure of the shape, and thus can be non-distinguishing for similar structures.


\section{Similarity Metrics}
\label{sec:matching}

The degree to which two shape descriptors match~\cite{veltkamp01} is quantified by a scalar similarity metric $M(\textbf{S}_i, \textbf{S}_j)$.  Since shape descriptors are vectors by construction, standard vector operations such as the Euclidean distance or vector projection provide natural similarity metrics.  Along this line, two standard similarity metrics, $\textbf{S}_i - \textbf{S}_j$ and $ \textbf{S}_i \cdot \textbf{S}_j $,  are defined.   The similarity metric based on the Euclidean distance is given by:
\begin{equation}
\textbf{S}_i - \textbf{S}_j = \left[ \sum_k (S_{i, k} - S_{j, k})^2 \right]^{1/2}.
\end{equation}
Here, $k$ is one component of the shape vector $\textbf{S}$, which may be a real or complex number.  Similarly, a similarity metric based on the projection of one shape descriptor vector onto another is defined by
\begin{equation}
\textbf{S}_i \cdot \textbf{S}_j = \left[ \sum_k (S_{i, k} S_{j, k}^*) \right]^{1/2}.
\end{equation}


For the sake of comparison, it is useful to define the similarity metrics on the interval $M \in [0, 1]$, with $1 (0)$ giving the maximum (minimum) match.  Thus, we redefine the Euclidean distance similarity metric as:
\begin{equation}
M_{dist}(\textbf{S}_i, \textbf{S}_j) = 1 - \left[ \left( \textbf{S}_{i} -  \textbf{S}_{j} \right) / \left( |\textbf{S}_i| + |\textbf{S}_j| \right) \right].
\end{equation}
Similarity, we redefine the projection-based similarity metric as:
\begin{equation}
M_{dot}( \textbf{S}_i, \textbf{S}_j ) = \dfrac{1}{2}\left[1 +  \left( \textbf{S}_{i} \cdot \textbf{S}_{j} \right) / \left( |\textbf{S}_i| |\textbf{S}_j | \right)  \right].
\end{equation}

The modified similarity metrics also have simple geometric interpretations.  The  $M_{dist}$ function is the ratio of the Euclidean distance between vectors and the maximal distance between the vectors (i.e. if the vectors are antiparallel).  The $M_{dot}$ function is proportional to the degree of spatial alignment between descriptor vectors.  If $\textbf{S}_i$ and $\textbf{S}_j$ are parallel, $M_{dot}$ has a value of 1.  If $\textbf{S}_i$ and $\textbf{S}_j$ are antiparallel, $M_{dot}$ has a value of 0.
After normalization, the only difference between similarity metrics is the proportional weight given to the two types of differences.  Matching functions based on projection are sensitive to differences in the signs of components, whereas distance-based metrics are only sensitive to do not magnitude of differences regardless of the sign.  

In addition to these metrics, we can define a wide variety of other metrics that are sensitive to particular differences in shape descriptors.  For existing metrics, differences or correlations can be dampened or accentuated by applying an arbitrary power $p$ to the component-wise comparison.  In some cases, highly specialized similarity metrics can be applied to specific descriptors.  An example of a specialized matching scheme is given by the quadratic metric of the shape histogram method of reference~\cite{osada}, which takes into account neighboring histogram bins when computing differences.

\section{Algorithms and Examples}
\label{sec:applications}


The shape descriptors and similarity metrics described in the previous sections can be used to create various types of order parameters, correlation functions, and other structural metrics.  In this section, we describe general algorithms that, when used with the appropriate shape descriptors and similarity metrics, can be applied to characterizing structure for a wide range of particle systems.  In some cases, the algorithms are reformulated versions of standard schemes from the condensed matter literature.  Additionally, we explore algorithms from the shape matching literature that have not yet been widely applied to particle systems, and present some completely new algorithms that exemplify the future direction of the framework.  For all of the algorithms, we provide representative example problems to demonstrate their application.  Our examples are mostly drawn from the self-assembly literature; however, in some cases we explore more idealized problems from the condensed matter literature for simplicity.  Since our goal is to present important elements of the shape matching framework rather than solve specific problems, the examples should be considered proofs-of-concept rather than optimal solutions.  

\subsection{Simple Structure Identification}
\label{ssec:identification}
The goal in most computer science shape matching applications is to identify unknown structures by searching a database of known reference structures.  Structures are identified by the known structure that gives the best match.  This type of scheme has already been applied to particle systems in the context of fast shape-based database searches for proteins and macromolecules~\cite{lesk86, ssm, proteinpmpnas, yeh, venkatraman, mak, grandison}.  The algorithm for structure identification is given in pseudocode below.
\begin{quote}
\begin{mylisting}
\begin{verbatim}
set match_best = 0
set id_best = `none'
call compute_shape_descriptor(S_i)
 
for each structure j in reference_database
   call compute_shape_descriptor(S_j)
   set match = M(S_i, S_j)
   if match > match_best
       match_best = match
       set id_best = id_j
   end if
end for
return id_best
\end{verbatim}
\end{mylisting}
\end{quote}
Although structure identification schemes based on database searches have been applied in limited cases for condensed matter systems~\cite{iac07, gyroid}, many standard structure identification schemes bear strong resemblance to this algorithm.  For example, the common neighbor analysis (CNA) scheme of reference~\cite{cna} involves identifying local clusters by matching their numerical fingerprints, based on their distribution of local neighbor configurations, with those for predetermined ideal structures.  In a rough sense, the CNA fingerprints can be considered shape descriptors for a given cluster, and the ideal fingerprints can be considered a database of reference structures.  A similar structural identification scheme is based on the bond order parameters of reference~\cite{snr83}.  In this scheme, local structures are identified by choosing cutoff values for various bond order parameters, beyond which query structures are said to match an ideal structure with a known high value of the order parameter~\cite{gasser}.  In this case, the bond order parameters represent shape descriptors, the cutoffs act as similarity metrics, and the ideal structures used to set the cutoffs act as the reference database.

\begin{figure}
\begin{center}
\includegraphics[width=0.85\columnwidth]{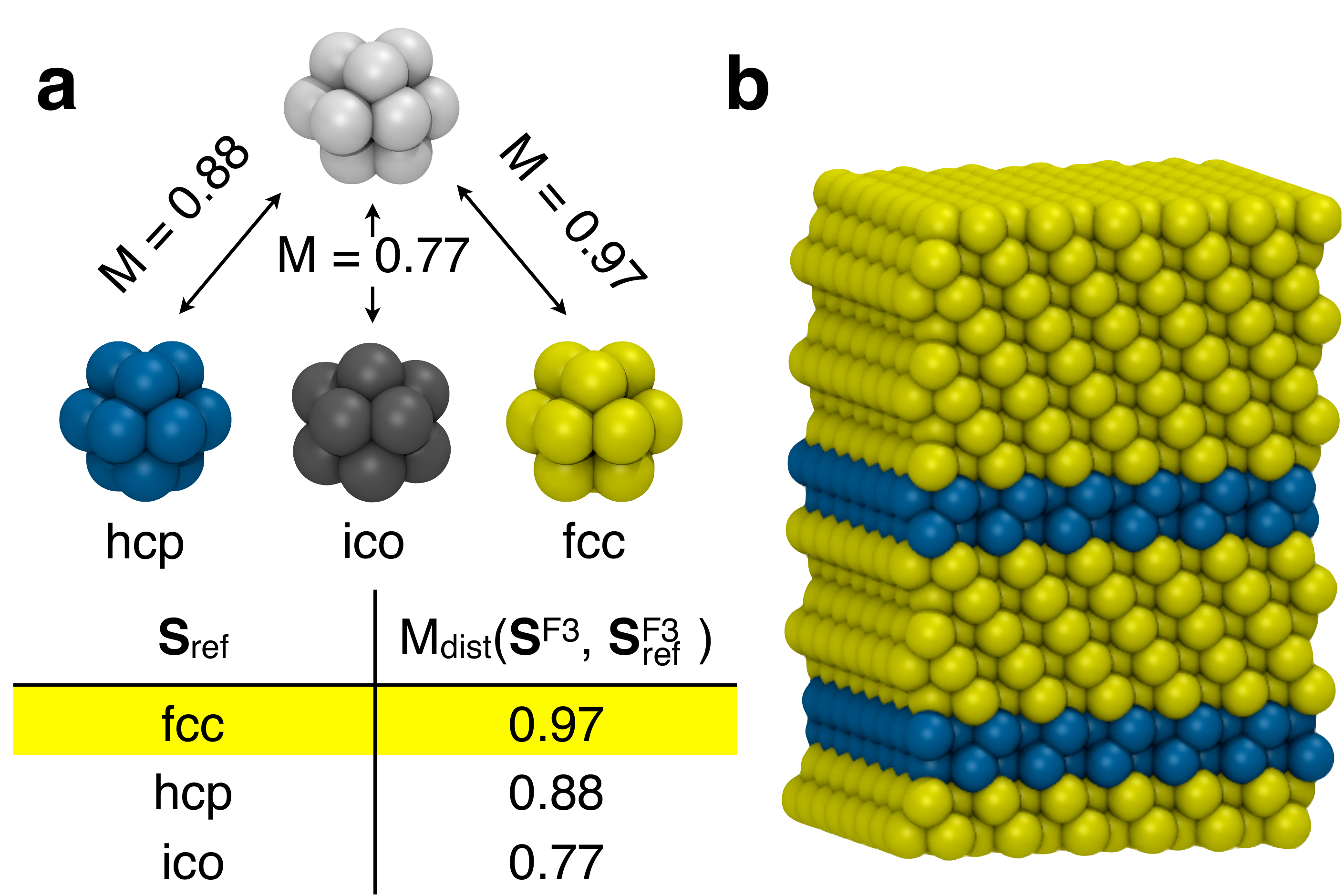}
\caption[Identification of Local Structures]{Identification of local structures  (\textit{a})  Basic identification of a slightly imperfect fcc cluster.  The table shows the matching values for the query structure compared to fcc, hcp and icosahedral reference clusters.  (\textit{b})  A fcc crystal with hcp stacking faults.  The particles are colored based on their first neighbor shell configuration.  Light (yellow) particles are in the fcc configuration, while dark (blue) particles are in the hcp configuration.}
\label{fig:stackingfault}
\end{center}
\end{figure}

As a minimal example of a structure identification scheme, consider the problem of identifying the small, imperfect cluster in Fig.~\ref{fig:stackingfault}a, where particles have been slightly perturbed from their ideal face-centered-cubic (fcc) positions.  For the purpose of the present example, we consider a small library of reference structures consisting of fcc, hexagonal-close-packed (hcp) and icosahedral clusters, each with 13 atoms.  To differentiate between these clusters, we use rotation invariant Fourier descriptors on the surface of the sphere $\textbf{S}^\textrm{F3}$, where $\ell = 4, 6$.  These coefficients are chosen because they are the leading coefficients for this class of structures~\cite{snr83}.  As shown in the table in Fig.~\ref{fig:stackingfault}a, the unknown structure best matches with the fcc cluster, followed by the hcp and icosahedral clusters, thus identifying the structure as fcc.

This simple matching scheme can be performed repeatedly to identify local structures in a global sample.  Consider the defective fcc crystal shown in Fig.~\ref{fig:stackingfault}b, which contains hcp stacking faults~\cite{tesfuv2}.  The stacking faults can be identified by finding particles in local hcp configurations rather than fcc.  First, local structural patterns must be created for each particle.  This is done by  clustering all neighboring particles within a cutoff radius $r_{cut}$.  Here, the cutoff is chosen to encompass the first peak in the radial distribution function $g(r)$; first neighbors can alternatively be found by the Voronoi construction~\cite{voronoi}.  Since we know \emph{a priori} that ideal fcc and hcp clusters are the only possible structures, our reference library consists of these two structures exclusively.  Particles are identified by finding a best match, as in the previous example, and colored based on their local configuration (light corresponds to fcc, dark to hcp) in Fig. \ref{fig:stackingfault}b, highlighting the stacking faults.

\begin{figure}
\begin{center}
\includegraphics[width=0.85\columnwidth]{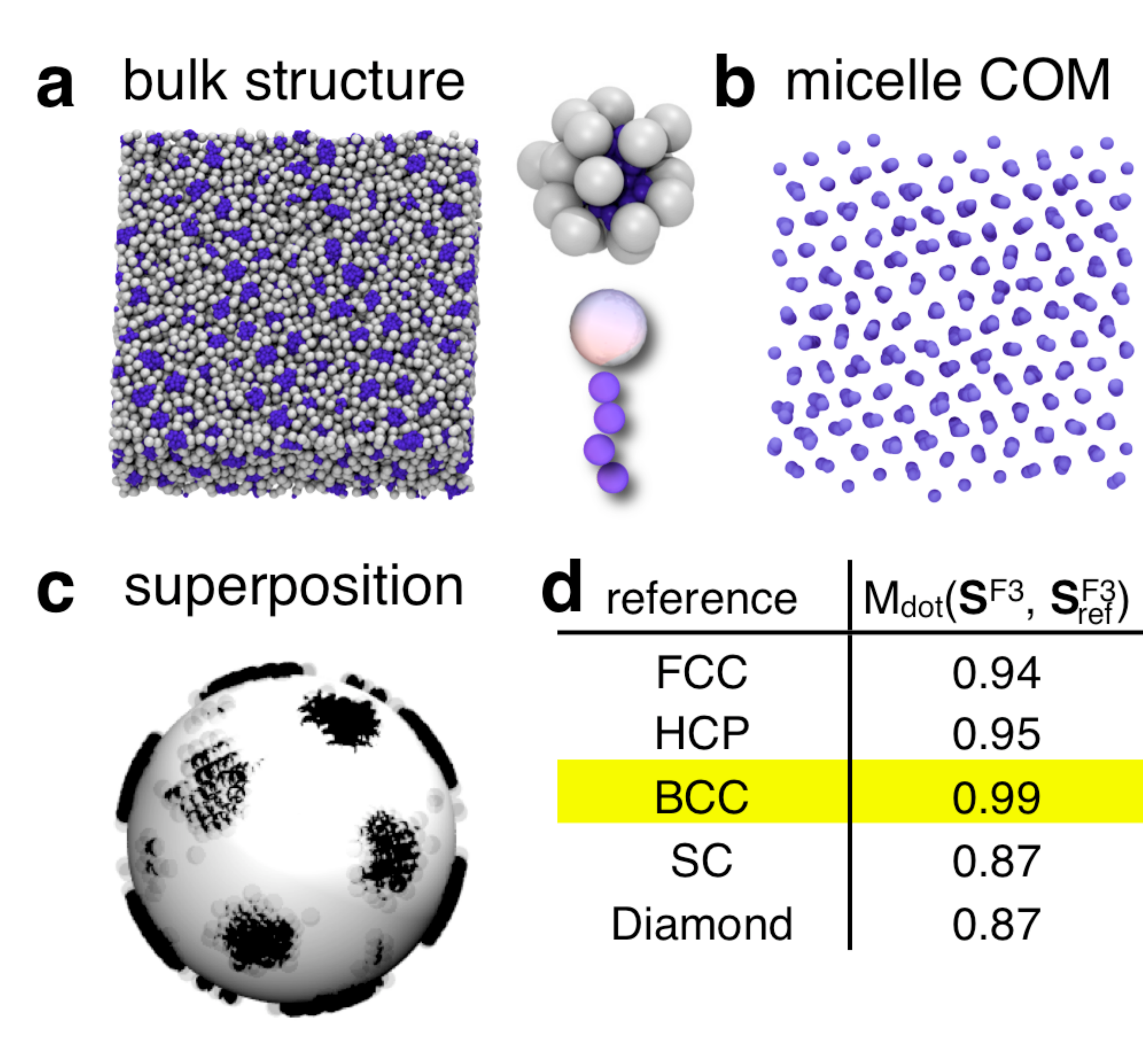}
\caption[Identification of Global Crystalline Structures]{Identification of global crystalline structures for a system of mono-tethered nanospheres that aggregate into spherical micelles. (\textit{a}) Bulk micelle structure. (\textit{b}) The micelle centers-of-mass are extracted using a Gaussian filter.  (\textit{c}) The global pattern is created by superposition of the local patterns. (\textit{d}) Matching is based on a Fourier descriptor (section~\ref{ssec:fourier}) that indexes the global superposition of local patterns (section~\ref{ssec:global}), and identifies the micelles as bcc structured.} 
\label{fig:micelles}
\end{center}
\end{figure}

Structure identification can also be performed for global samples.  Global structure identification can be useful, e.g., when mapping structural phase diagrams (see section~\ref{ssec:spatialcorrelations}).  As an example of global structure identification, consider the mono-tethered nanosphere system similar to references~\cite{iacovella2005, zhang2003}, whose tethers phase separate into spherical micelles (see Fig.~\ref{fig:micelles}a).  The micelles themselves pack into an ordered crystalline superstructure.  The structure of the crystals can be determined by identifying the micelle centers of mass, which make up the set of positions $\{X\}$ that describe the system  (see Fig.~\ref{fig:micelles}b).  The centers of mass are determined by applying a Gaussian filtering algorithm adapted from the colloidal science literature~\cite{varadan2003, crocker1996}.  A global crystalline pattern is determined by computing the superposition of local patterns (see Fig.~\ref{fig:micelles}c, similar to Fig.~\ref{fig:slglobal}a).  The global pattern is then compared to that for several standard candidate crystals, by matching Fourier descriptors for patterns on the surface of the sphere: $M_{dot}(\textbf{S}^\mathrm{F3}, \textbf{S}^\mathrm{F3}_{ref})$.  Here, the Fourier descriptor is composed of the leading terms in the harmonic expansion for the standard crystals: $\textbf{S}^\mathrm{F3} = < |\textbf{q}_{4}|, |\textbf{q}_{6}|, |\textbf{q}_{8}|, |\textbf{q}_{10}|, |\textbf{q}_{12}| > $.  Notice that we use invariant Fourier coefficients for rotation-independent matching.  The unknown crystal is identified by the reference structure that gives the best match, in this case bcc (see Fig.~\ref{fig:micelles}d).

The structure identification applications presented in this section are successful because the potential reference structures are known \textit{a priori.}  However, the identification schemes fail if the unknown structure is not in the reference database.  It is therefore important to carefully choose the appropriate reference structures for a given application.  Often, optimal matches are obtained by using imperfect structures from the system rather than mathematically perfect structures for reference structures.  As an added consideration, it is sometimes possible to obtain partial structures that are highly ordered, but are missing one or more particles.  This is an important factor in phase separated systems, systems with physical boundaries, and systems with high variation in neighbor distances.  For proper identification, partial structures must be added to the reference library explicitly~\cite{iac07}, unless a shape descriptor capable of partial matching is used, such as a point matching descriptor or shape contexts~\cite{belongie}.

\subsection{Identification Without Reference Structures}

As the number of potential structures grows, compiling a comprehensive reference library becomes increasingly difficult.  However, if the space of potential reference structures is finite, a reference library can be created on-the-fly (OTF), precluding the need to define reference structures \textit{a priori.}  To do so, a new reference structure is added to the library whenever no suitable match is found.   The algorithm for identification without known reference structures is given in pseudocode below.
\begin{quote}
\begin{mylisting}
\begin{verbatim}
set match_best = 0
set id_best = `none'
call compute_shape_descriptor(S_i)
 
for each structure j in reference_database
   call compute_shape_descriptor(S_j)
   set match = M(S_i, S_j)
   if match > match_best
       match_best = match
       set id_best = id_j
   end if
   if match_best < match_min
      call add_structure_to_database(S_i, counter)
      set id_best = counter
      set counter = counter + 1
   end if
end for
return id_best
\end{verbatim}
\end{mylisting}
\end{quote}
Notice that the algorithm requires an additional step where the reference structures themselves, which are initially ``unnamed,'' are identified.  This can be accomplished by using a standard identification algorithm similar to that outlined in section~\ref{ssec:identification}, or in some cases more simply by visual inspection.


\begin{figure}
\begin{center}
\includegraphics[width=1.0\columnwidth]{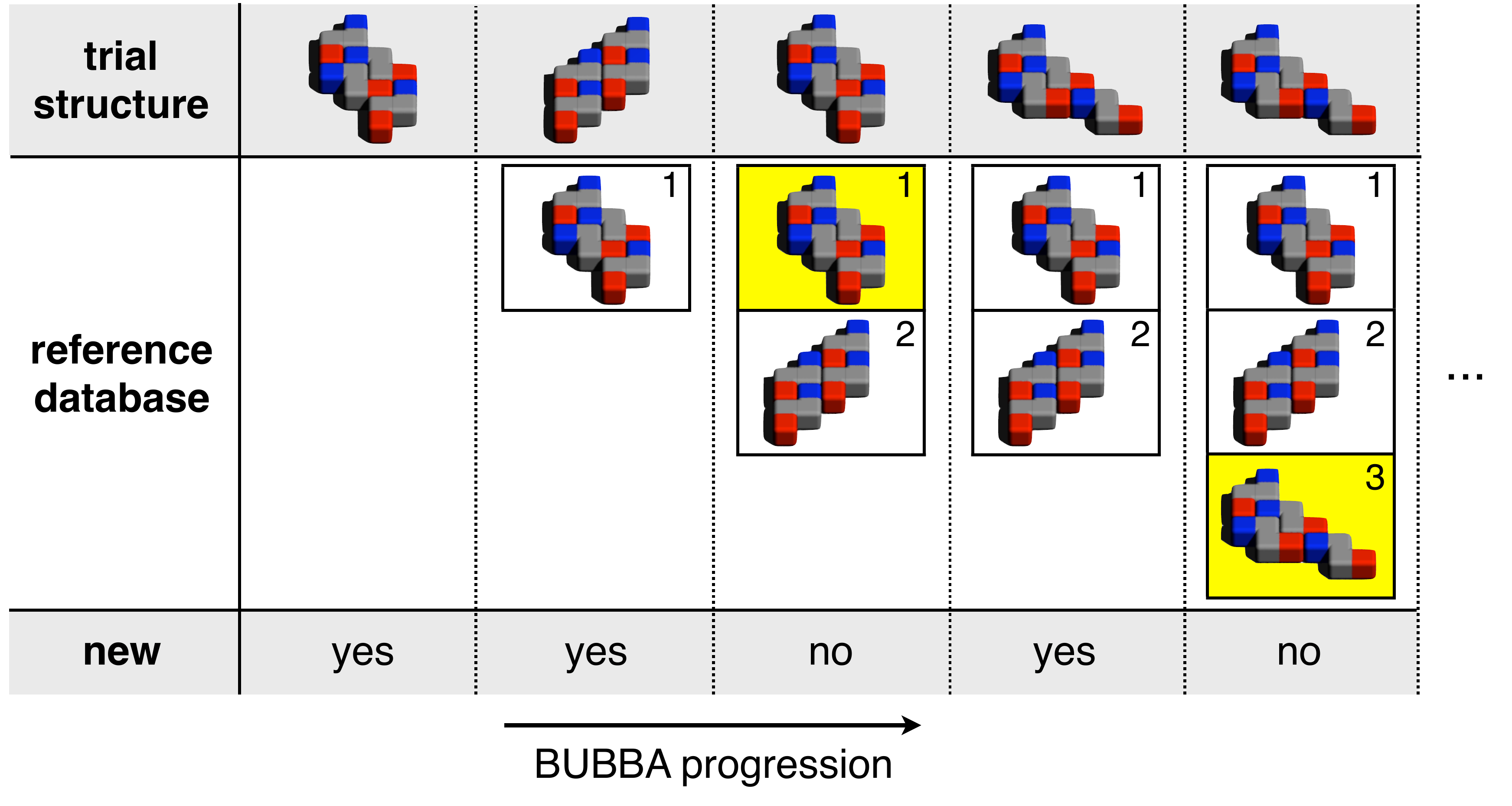}
\caption[On-the-fly Reference Library with Bottom Up Building Block Assembly (BUBBA)]{On-the-fly reference library with Bottom Up Building Block Assembly (BUBBA)~\cite{bubba}.  The BUBBA algorithm involves enumerating unique clusters of a given size $N$.  To ensure that clusters are unique, new clusters are added to the reference library and given a unique identifier, while repeated clusters (yellow) are discarded.} 
\label{fig:bubba}
\end{center}
\end{figure}

The OTF algorithm is applicable to simulations or algorithms that involve enumerating unique structures.  One example is given by Bottom-Up Building Block Assembly (BUBBA)~\cite{bubba}.  The BUBBA algorithm efficiently generates low-energy clusters by trying different combinations of smaller low-energy clusters.  To ensure that the clusters generated are not redundant, an OTF shape matching scheme is employed (see Fig.~\ref{fig:bubba}).  New clusters are added to a reference library if no match is found, while redundant clusters are ignored.  In the end, the reference library contains a list of unique clusters. This type of scheme may also potentially be applied to information compression for mapping structural phase spaces.  Since large portions of the parameter space are often redundant for high-resolution mappings, an OTF scheme can be employed to quickly obtain a minimal number of unique structures.

\subsection{Identification in Systems with Disordered Structures}
Creating a comprehensive reference library is nearly impossible when local structures can assume disordered configurations.  In this case, the space of potential reference structures is essentially infinite, since ``disordered'' refers to the vast space of configurations with no particular structure.  As a solution, a structure that does not match any structure in the reference library within a certain threshold is considered ``disordered ~\cite{iac07}.''  This requires that we choose a cutoff value for a best match.  The cutoff must be chosen carefully; in thermal systems, an overly-stringent cutoff might cause a matching scheme to miss highly-ordered structures perturbed slightly from their ideal configurations, whereas an overly-permissive cutoff can misidentify highly disordered structures.  In most cases, a sufficiently rigorous cutoff can be defined such that its value does not affect the qualitative results.  The algorithm for structure identification with disordered local structures is given in pseudocode below:
\begin{quote}
\begin{mylisting}
\begin{verbatim}
set match_best = 0
set id_best = `none'
call compute_shape_descriptor(S_i)
 
for each structure j in reference_database
   call compute_shape_descriptor(S_j)
   set match = M(S_i, S_j)
   if match > match_best
       match_best = match
       set id_best = id_j
   end if
   if match_best < disordered_cut
      id_best = `disordered'
   end if
end for
return id_best
\end{verbatim}
\end{mylisting}
\end{quote}

\begin{figure}
\begin{center}
\includegraphics[width=0.9\columnwidth]{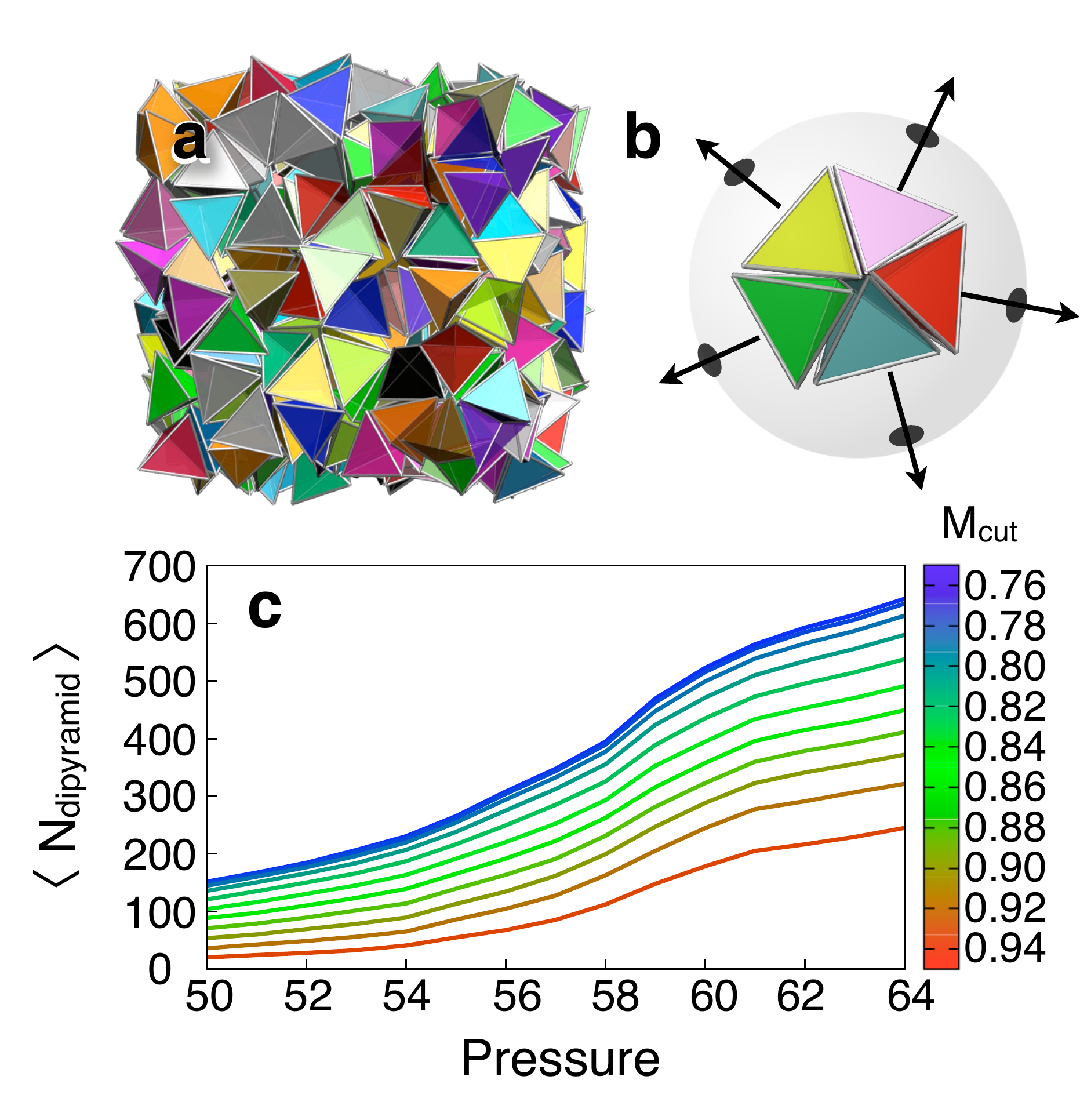}
\caption[Pentagonal Dipyramids (PDs) and Transitions the Hard Tetrahedron System]{Pentagonal dipyramids (PDs) in the hard tetrahedron system~\cite{amir09}. (\textit{a}) A snapshot of the hard tetrahedron liquid at packing density $\phi \approx 0.5$ and reduced pressure $P=60$. (\textit{b}) A PD-like cluster taken from the system. The arrows depict the pattern of directions $[ \theta, \phi ]$ on the surface of the sphere indexed for matching. (\textit{c}) The number of PDs as a function of the identification cutoff value. Notice that for all cutoffs, there is an inflection point centered at $P=58$, which corresponds to a possible liquid-liquid transition marked by a sudden increase in PD-like local ordering.}
\label{fig:tetrahedra}
\end{center}
\end{figure}

As an example, consider the hard tetrahedron fluid studied in reference ~\cite{amir09} (Fig.~\ref{fig:tetrahedra}a).  In this system, an important local motif, originally identified by visual inspection, is the ``pentagonal dipyramid'' (PD), formed by five tetrahedra sharing a common edge.  The PDs form a spanning network as the system goes through a liquid-liquid transition.  To identify PDs, we first cluster all sets of five tetrahedra in the system that form a closed polygon.  The shape of each cluster is defined by projecting the directions of the tetrahedra on the surface of a sphere (Fig.~\ref{fig:tetrahedra}b).  An ideal PD gives a 2D pentagonal pattern, which is taken as a reference structure.  Although the pentagon reference structure is confined to a plane, the query structures are not.  Therefore, this pattern is well-described by Fourier descriptors on the surface of the sphere, with matching given by  $M_{dist}(\textbf{S}^\mathrm{F3}_{query}, \textbf{S}^\mathrm{F3}_{pentagon})$.  We take rotation-invariant descriptors with frequency parameter $\ell=5,6,...10$.  For some systems, there is a clear distinction between ordered and disordered structures.  However, for the tetrahedron system we observe a continuous spectrum of PD-like ordering.  Thus, we estimate a cutoff based on visual inspection in the range $M_{cut} \sim 0.9$.  Although this choice is arbitrary, it has little effect on structural trends for the system.  Fig.~\ref{fig:tetrahedra}c shows the number of PDs for a wide range of cutoffs.  We see that for all cutoffs, the fraction of PDs exhibits a weak crossover, marked by an inflection point, near reduced pressure $P=58$.  This pressure, in turn, corresponds to an interesting thermodynamic transition for the system~\cite{amir09}.  Although cutoffs result in different numbers of PDs, the same underlying physical behavior is captured regardless.

\subsection{Order Parameters and Temporal Correlation Functions}
Another standard application of structural metrics is to track structural transitions, either as a function of time or a changing reaction coordinate.  This is typically accomplished by monitoring either an order parameter or correlation function as the system changes.  In the context of the shape matching framework, the difference between the two cases is largely semantic; while an order parameter typically measures similarity with an ideal structure, a correlation function typically measures similarity between different structures in the system, separated in time and/or space.   The simple algorithm for tracking a transition as a function of a changing parameter is given below:
\begin{quote}
\begin{mylisting}
\begin{verbatim}
call compute_shape_descriptor(S_ref)
for p in changing_parameter
   call compute_shape_descriptor(S_p)
   set order_param[p] = M(S_p, S_ref)
end for
return order_param
\end{verbatim}
\end{mylisting}
\end{quote}
Here the similarity metric $M \in [0, 1]$ serves as a convenient order parameter. Tracking structural transitions is important for a wide variety of applications, including elucidating thermodynamic transitions ~\cite{tenwolde96,  halperin74, smectic, gubbins, confinedfluids} and assembly pathways~\cite{zhenlidiamond, yamaki, bladon93, tang2006}.  Many of the advanced molecular simulation techniques used to study transitions~\cite{torrie77, tps, bolhuis2002transition, ffs, metadynamics} rely on structural metrics in the context of pseudo-reaction coordinates~\cite{tps}, biasing parameters~\cite{torrie77}, and collective variables~\cite{metadynamics} to guide the statistical sampling algorithm.  Standard order parameters have been devised for various types of ordering, including bond orientational ordering~\cite{halperin78, nelsonc6, snr83, steinhardt81}, liquid crystalline ordering~\cite{liquidcrystals, larson} such as nematic~\cite{nematic} and smectic~\cite{smectic} phases, chiral ordering~\cite{kamienchiral}, and helical ordering~\cite{h4}.  Time correlation functions based on these order parameters have been applied to creating structural autocorrelation, or ``memory'' functions for glassy liquids~\cite{kawasaki, tanaka} and growing quasicrystals~\cite{keys07}. 

\begin{figure}
\begin{center}
\includegraphics[width=0.95\columnwidth]{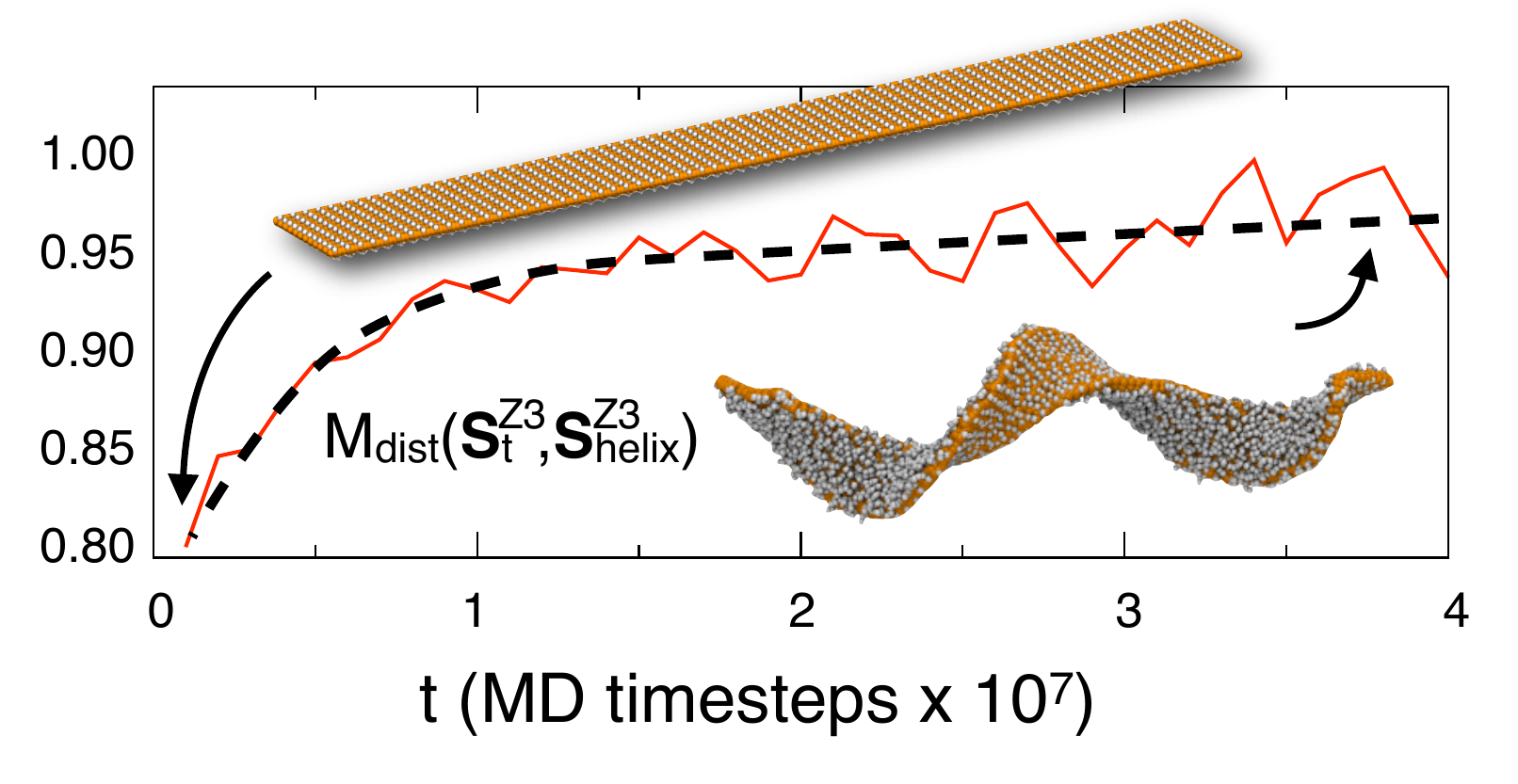}
\caption[Assembly of a Helical Sheet Composed of Laterally Tethered Nano-Rods]{Assembly of a helical sheet composed of laterally tethered nano-rods~\cite{trunghelix}.  As time progresses, the initially flat sheet twists into a helix.  The matching order parameter $M_{dist}(\textbf{S}_t, \textbf{S}_{helix})$ compares the structure at time $t$ with the shape of the final ideal helical structure.}
\label{fig:slhelix}
\end{center}
\end{figure}

As a simple example of creating an order parameter within the shape matching framework, consider the sheet-like structure self-assembled from laterally tethered nano-rods studied in reference~\cite{trunghelix}, and shown in Fig.~\ref{fig:slhelix}a.  Due to an instability, the initial sheet relaxes into a helical structure that minimizes the free energy.  We can track this structural transition by matching the shape of the sheet at a given time $t$ with the final, fully equilibrated helical structure: $M(\textbf{S}_t, \textbf{S}_{helix})$.  Since the structure is 3-dimensional and has radial dependence, it can be indexed using a Zernike descriptor on the unit ball, $\textbf{S}^\mathrm{Z3}$.  Since the sheet only changes in terms of its twist in space, we save computational effort by only considering points along the backbone of the sheet.  To match the shape independently of the orientation of the sheet, we take rotation invariant moments with $\ell$ in the range $4 \leq \ell \leq12$.  Fig \ref{fig:slhelix}a shows the helical order parameter as a function of time for a long molecular dynamics run.  We observe that over tens of millions of MD steps, the sheet slowly and continuously equilibrates to the final helical structure, in agreement with visual inspection.  Our matching order parameter gives a better indication of the structural transition than the more standard helical order parameter $H4$~\cite{h4}, which is rather insensitive when the pitch of the helix is large compared to the radius~\cite{trunghelix}.  The noise in the data at long $t$ is indicative of the relatively large fluctuations in shape that occur in equilibrium.  


\begin{figure}
\begin{center}
\includegraphics[width=0.9\columnwidth]{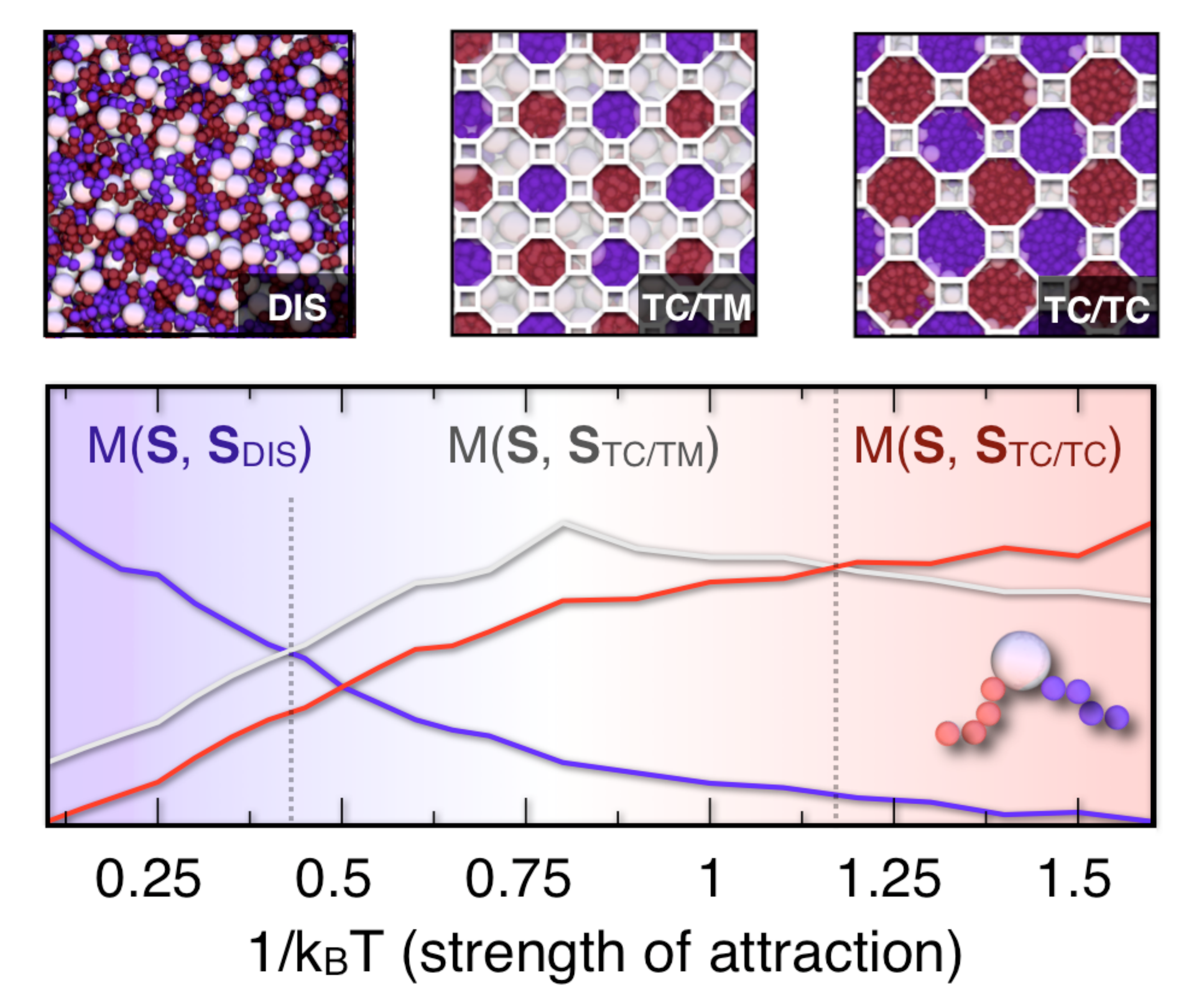}
\caption[Structural Transitions Phase Separated System ]{Structural transitions in a phase separated system. (\textit{a}) Visual depiction of the three structures formed by a ditethered nanosphere system~\cite{iacovella2009b} (left to right: disordered, TC/TM, TC/TC). (\textit{b}) Matching order parameter for the three reference structures as a function of inverse temperature.}
\label{fig:slditns}
\end{center}
\end{figure}

As a slightly more complex example, consider the structures formed by the ditethered nanospheres shown in Fig.~\ref{fig:slditns}a~\cite{iacovella2009b}.  The system goes through two transitions as a function of inverse temperature or quench depth, first from a disordered structure to a phase-separated structure characterized as a tetragonal cylinder/tetragonal-mesh (TC/TM), and then to a similar structure characterized by tetragonal cylinders (TC/TC)~\cite{iacovella2009b}.   The abbreviations indicate the patterns formed by the tethers and nanoparticles, respectively.  To obtain a quantitative measure of this behavior, we take three reference points: ideal snapshots from the disordered regime, the TC/TM regime and the TC/TC regime.  As outlined in section~\ref{ssec:global}, several different descriptors are applicable to this type of global structure.  Here, we use the shape distribution $\textbf{S}^{\mathrm{D2}}$, since it is distinguishing between the three compared structures.  Separate $\textbf{S}^{\mathrm{D2}}$ descriptors are created for each of the three aggregating species.  These descriptors are then concatenated into an overall descriptor.  Rather than considering surface particles exclusively, we use all particle positions, since this is simpler and is still distinguishing for the cylindrical phases under consideration.  Fig.~\ref{fig:slditns}b shows how the character of the system changes as a function of inverse temperature.  We see that the structural transition between the three phases is smooth and continuous, as verified by visual inspection.  In a separate reference~\cite{keys10-HO}, we show that a scheme based on the distribution of Fourier descriptors for local density maps gives identical results.

\begin{figure}
\begin{center}
\includegraphics[width=0.8\columnwidth]{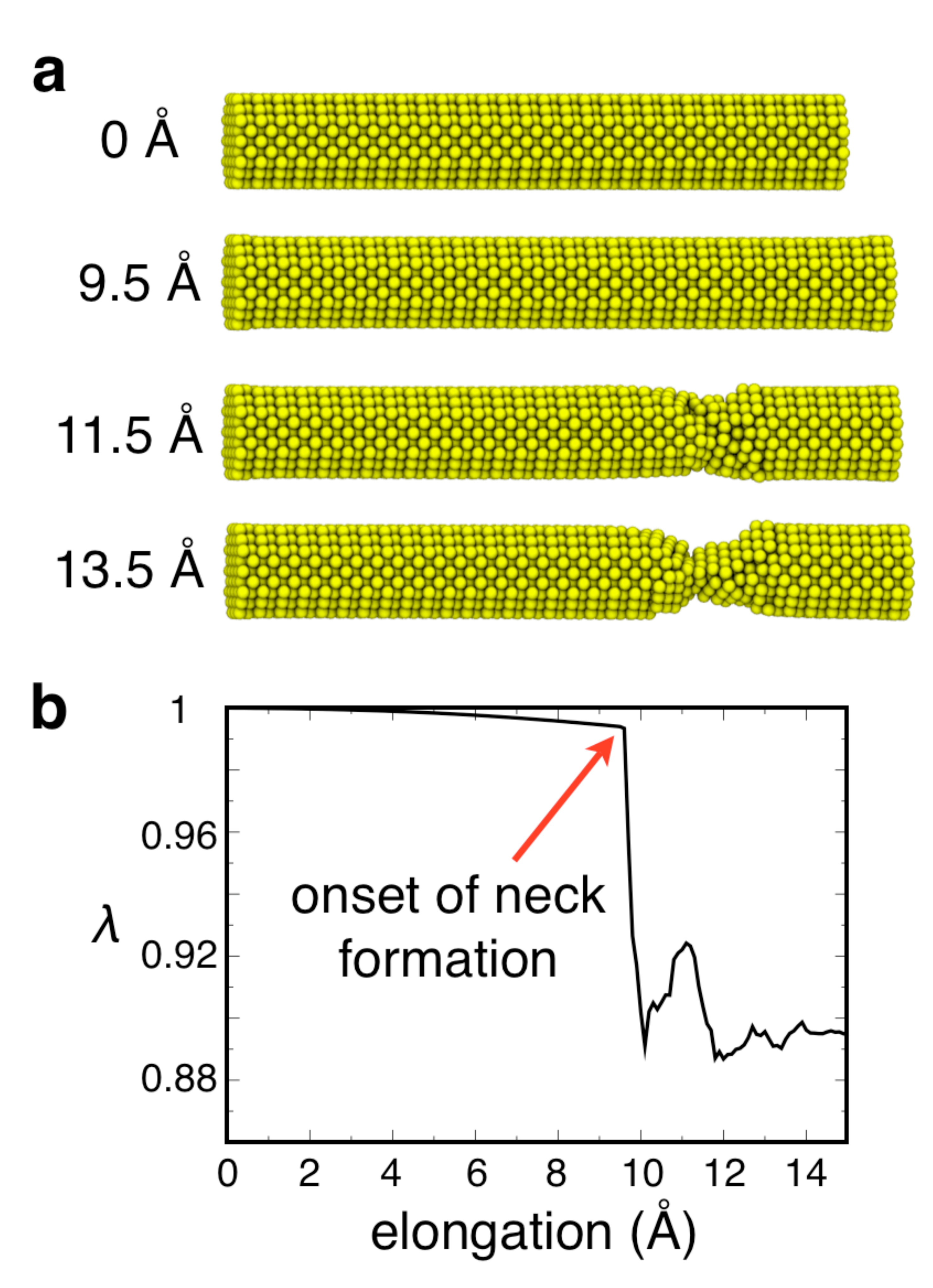}
\caption[Neck Formation in a Gold Nanowire]{Neck formation in a gold nanowire~\cite{Pu2008, iacovella2010}. (\textit{a}) depiction of the nanowire as a function of elongation $L$.  (\textit{b}) The standard deviation of matching values, $\lambda = 1-\left< M_{dot}(\textbf{S}^\mathrm{F3}_{L}, \textbf{S}^\mathrm{F3}_{L=0}) \right>_{stdev}$ as a function of elongation.}
\label{fig:nanowire}
\end{center}
\end{figure}

As a relatively complex example, consider the gold nanowire undergoing tensile elongation shown in Fig. \ref{fig:nanowire}a ~\cite{Pu2008,iacovella2010}.  As the wire elongates, a ``neck'' begins to form, in this case, at $\sim$10 $\AA$.  This type of structural transition can strongly impact the transport properties of nanowires~\cite{Barnett1997} and thus is important to identify.  The neck region can be characterized by the loss of the original fcc structure locally, e.g. a change in orientation, number of neighbors, or overall symmetry.  Global crystalline order parameters~\cite{snr83} may not be well suited, since only a subset of the system undergoes a transition.  Standard schemes that differentiate between crystal and liquid configurations locally, such as $\textbf{q}_6 \cdot \textbf{q}_6$~\cite{tenwolde96} (see section~\ref{ssec:spatialcorrelations}), are not well suited either, since the finite nature of the nanowire results in neighboring atoms having different local coordinations, even in the ideal fcc configuration (i.e. a mixture of full and partially coordinated fcc clusters).  Instead, to detect the onset of necking, we compare an atom's neighbor shell to its initial structure as a function of elongation $L$: $M(\textbf{S}_L, \textbf{S}_{L=0}$).  Local neighbor shells are indexed using rotation-dependent Fourier descriptors $\textbf{S}^\mathrm{F3} = \textbf{q}_6$, where $\ell=6$ is chosen because it has been shown to describe fcc clusters well without requiring other frequencies~\cite{tenwolde96, auer04}.   The number of atoms in the neck is small compared to the bulk, thus the average autocorrelation value is not strongly sensitive to neck formation.  However, the spread in the data is sensitive to neck formation, and rapidly increases when atoms in the neck lose their original structure and yield low matching values.  We can therefore create an \textit{ad hoc} order parameter based on $\lambda = 1- \left< M_{dot}(S^{F3}_{L}, S^{F3}_{L=0}) \right>_{stdev}$, where $\lambda=1$ for the ideal configuration at $L=0$, and decreases proportionally as the spread in matching values increases.  Fig.~\ref{fig:nanowire}b shows that the onset of neck formation occurs at 9.7 $\AA$, which is consistent with visual inspection. 


\subsection{Spatial Correlation Functions}
\label{ssec:spatialcorrelations}

\begin{figure*}
\begin{center}
\includegraphics[width=0.65\textwidth]{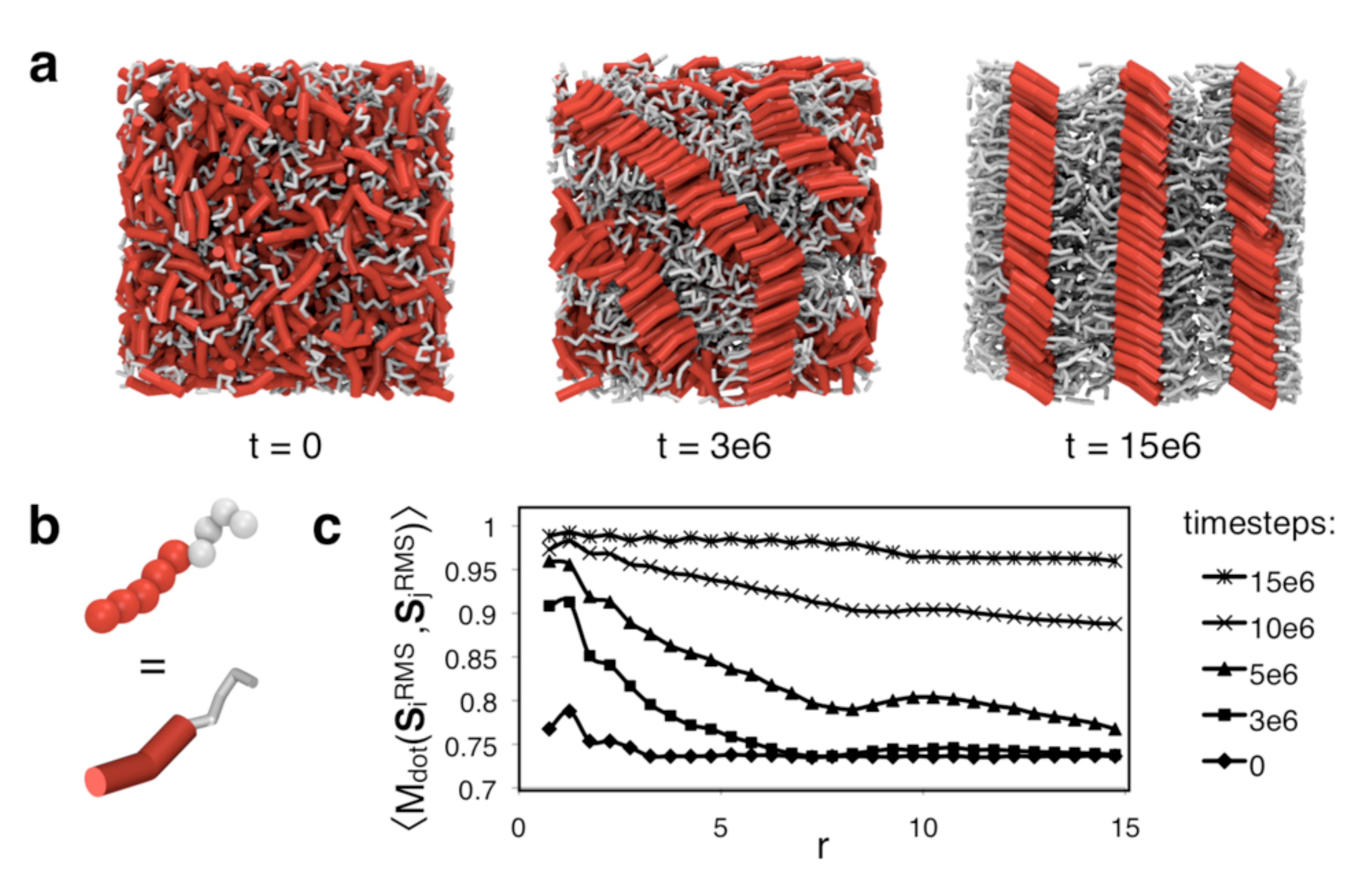}
\caption[Spatial Correlations in a System of  Tethered Nano ``V's'']{Spatial correlations in a system of  tethered nano ``V's''~\cite{tnv}.  (\textit{a})~Depiction of the formation of a lamellar phase as the system evolves in time on cooling.  (\textit{b})~Depiction of the coarse-grained nanoparticle model.  (\textit{c})~Nanoparticle orientational correlations as a function of separation distance $r$.}
\label{fig:tnv}
\end{center}
\end{figure*}


In addition to characterizing how structures change as a function of time or a reaction coordinate, another common application of structural metrics is to characterize how structures change in space.   In the context of the shape matching framework, this involves choosing structures from different points in the system, rather than ideal structures, as reference structures. Spatial correlation functions are often used to measure structural ``correlation lengths.''   The algorithm for computing structural correlation lengths within the matching framework is given in pseudocode below:
\begin{quote}
\begin{mylisting}
\begin{verbatim}
for i in list_of_local_structures
   call compute_shape_descriptor(S_i)
   for j in list_of_local_structures
       call compute_shape_descriptor(S_j)
       set r =  distance(i, j)
       set correlation_function(r) += M(S_i, S_j)
       set normalization(r) = normalization(r) + 1
end for
return correlation_function / normalization
\end{verbatim}
\end{mylisting}
\end{quote}
In the condensed matter literature, structural correlation functions have been defined for crystal-like ordering in 2d~\cite{halperin78, nelsonc6}, and 3d~\cite{snr83, tenwolde96}, nematic ordering~\cite{allen97}, and many other more specialized types of ordering.  Other types of spatial correlation functions have been widely applied as well.  One example is the $\textbf{q}_6 \cdot \textbf{q}_6$ scheme of references~\cite{tenwolde96, auer04}, which detects ordered crystal nuclei based on spatial correlations between local bond order parameters.

As a simple example of creating a spatial correlation function within the shape matching framework, consider the problem of characterizing the formation of  lamellae (sheets) in the system depicted in Fig.~\ref{fig:tnv}a, composed of tethered V-shaped nanoparticles (Fig.~\ref{fig:tnv}b)~\cite{tnv}.  From visual inspection, it is clear that the nanoparticles have long-range orientational correlations in the lamellar phase, but not in the disordered phase.  This can be quantified by computing an orientational correlation function for nanoparticles as a function of separation distance $r$.  It may initially appear that the nematic descriptor can be applied to this problem; however, since the nanoparticles have two directors, we lose important information about particle packing by considering only one angle.  Rather, an optimal metric reflects the correspondence between both directors of the nanoparticles.  This can be measured by a scheme based on the RMS descriptor, where the datapoints $\{X\}$ for each nanoparticle are given by the two directors, pointing from the vertex $\textbf{x}_v$ to each of the endpoints $\textbf{x}_{1}$, $\textbf{x}_{2}$: $\{X\} = \{\textbf{x}_{1} - \textbf{x}_v, \textbf{x}_{2} - \textbf{x}_v \}$.  Assignment of corresponding vectors is performed using the ``naive'' method (see section~\ref{ssec:RMS}), which is exact for this particular problem.  Fig.~\ref{fig:tnv}b shows the average value of the orientational correlation function $\left< M_{dot}(\textbf{S}_i^{RMS}, \textbf{S}_j^{RMS} (r) \right>$ as a function of the radial separation $r$ for several different snapshots.  In the disordered phase, only very short range correlations are present and $\left< M_{dot}(\textbf{S}_i^{RMS}, \textbf{S}_j^{RMS} (r) \right>$ is small for all $r$.  The correlations quickly grow as the system begins to form sheets, and the range of $\left< M_{dot}(\textbf{S}_i^{RMS}, \textbf{S}_j^{RMS} (r) \right>$ increases.  In the final state, the lengthscale is infinite, spanning the length of the simulation cell.  


\begin{figure}
\begin{center}
\includegraphics[width=0.7\columnwidth]{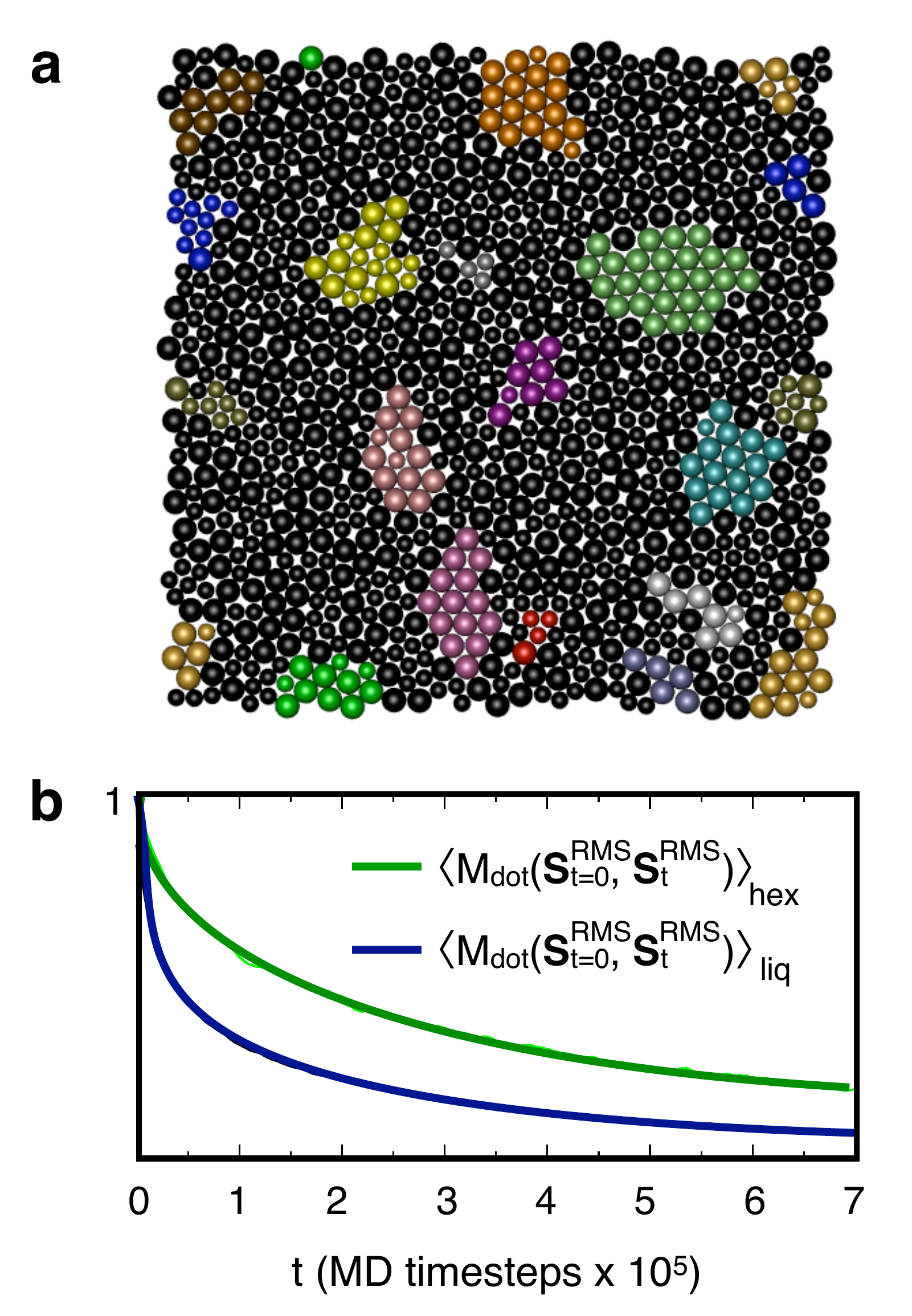}
\caption[Spatial and Temporal Correlations in Hexagonal Clusters] {Spatial and temporal correlations in hexagonal clusters.  \textit{a}) Depiction of the 2d hexagonal grains identified in the system.  Each grain is given a different color while liquid-like particles are colored black.  \textit{b})  Time-dependent structural decorrelation function for the hexagonal grains and the overall liquid.  Notice that the hexagonal grains retain their structure much longer than the overall liquid.}
\label{fig:hexclusters}
\end{center}
\end{figure}

As a slightly more complex example, consider the problem of measuring time-dependent structural correlations in the 2D binary mixture shown in Fig.~\ref{fig:hexclusters}a.  The system consists of a 50:50 mixture of spherical particles with a diameter ratio 1.4:1, and can represent either a model supercooled liquid~\cite{perera99, widmer2004reproducible} or a granular system near the onset of jamming~\cite{abate2006approach, keys2007measurement}.  The system contains small hexagonal crystal (hex) grains arranged randomly within the disordered bulk liquid.  To measure the effect of hex structure on dynamics, we can compare the rate of structural decorrelation in the hex and non-hex regions.  To do so, we require two correlation functions: first, a spatial correlation function to identify the hex regions and, second, a temporal correlation function to quantify how closely a given structure ``remembers'' its initial configuration as a function of time.  

Identifying the hex grains within the bulk liquid requires a structural criterion that differentiates between particles in the liquid and hex regions on a per-particle basis.  As mentioned previously, the $\textbf{q}_6 \cdot \textbf{q}_6$ scheme of reference~\cite{tenwolde96} can be used to monitor nucleation and growth in 3d systems that form fcc, bcc, and hcp crystals~\cite{auer04, tenwolde97, cacciuto04, gasser01, tesfuv2, laura}.  Although the scheme was originally based on the $\ell=6$ Fourier coefficient $\textbf{q}_6$, other shape descriptors can be just as easily be substituted~\cite{zhenlidiamond, keys07}.  The main physical insight underlying the $\textbf{q}_6 \cdot \textbf{q}_6$ scheme is that crystals often contain local particle configurations that match with their neighbors in terms of \textit{both} shape and orientation.  Therefore, crystal-like particles can be identified by detecting those that match well with their neighbors.  In analogy with reference~\cite{tenwolde96}, a good indicator of local crystal-like local ordering is the fraction of solid-like matches with neighbors:
\begin{equation}
f_{solid} = 1 / n \sum_j^n \Theta[M_{dot}(\textbf{S}_i,\textbf{S}_j) - M_{cut}].
\end{equation}
Here $\Theta$ is the Heaviside function and ``$j$'' is a neighbor of ``$i$,'' and \textbf{S} is a rotation-\textit{dependent} shape descriptor.  Particles with a minimal fraction of solid-like matches $f_{cut}$ are considered to be locally crystalline.  The cutoffs can be determined by viewing plots of the $P(M_{dot}(\textbf{S}_i,\textbf{S}_j))$ and $P(f_{solid})$ distributions for the bulk liquid and bulk solid~\cite{tenwolde96, auer04}, or simply by visual inspection, as we perform here.  This scheme holds for crystals in general, provided the neighbor shells all have the same shape.  In reference~\cite{keys10-HO}, we describe how this scheme can be modified to handle crystals with an assortment of neighbor shells.  The algorithm for detecting crystal grains is given in pseudocode below:
\begin{quote}
\begin{mylisting}
\begin{verbatim}
for i in list_of_particles
   set S[i] = compute_neighbors(i, rcut)
   call compute_rotation_dependent_shape_descriptor(S[i])
end for

for i in list_of_particles
   set n_solid = 0
   set count = 0
   for j within rcut of i
       if M(S[i], S[j]) > M_cut
          n_solid = n_solid + 1
       end if
       count = count + 1
    end for
    if n_solid / count > f_cut
        append(list_of_solid_particles, i)
    end if
end for
\end{verbatim}
\end{mylisting}
\end{quote}
%

 
For our example, we identify hex grains using the $\textbf{S}^\textrm{RMS}$ descriptor with $M_{cut} = 0.99$ and $ f_{solid} = 0.5$.  Notice that this scheme does not simply detect local hexagons; rather it identifies local hex regions while making the important distinction between isolated hexagons and the intermediate-range hex clusters of interest.  Our temporal correlation function is defined by matching the point-matching descriptor for each cluster at time $t$ with itself at a reference time $t_0$:  $ M_{dot}[\textbf{S}^\textrm{RMS}(t),\textbf{S}^\textrm{RMS}(t_0)] $.  Fig.~\ref{fig:hexclusters}b shows the correlation function for hexagonal and non-hexagonal particles.  We see that the hexagonal particles retain their structure longer on average than non-hexagonal particles.  Similar correlation functions have been applied to study glassy dynamics~\cite{kawasaki} and to study how the structure of liquid clusters change as they attach to a growing quasicrystal ~\cite{keys07}.

\begin{figure}
\begin{center}
\includegraphics[width=1.0\columnwidth]{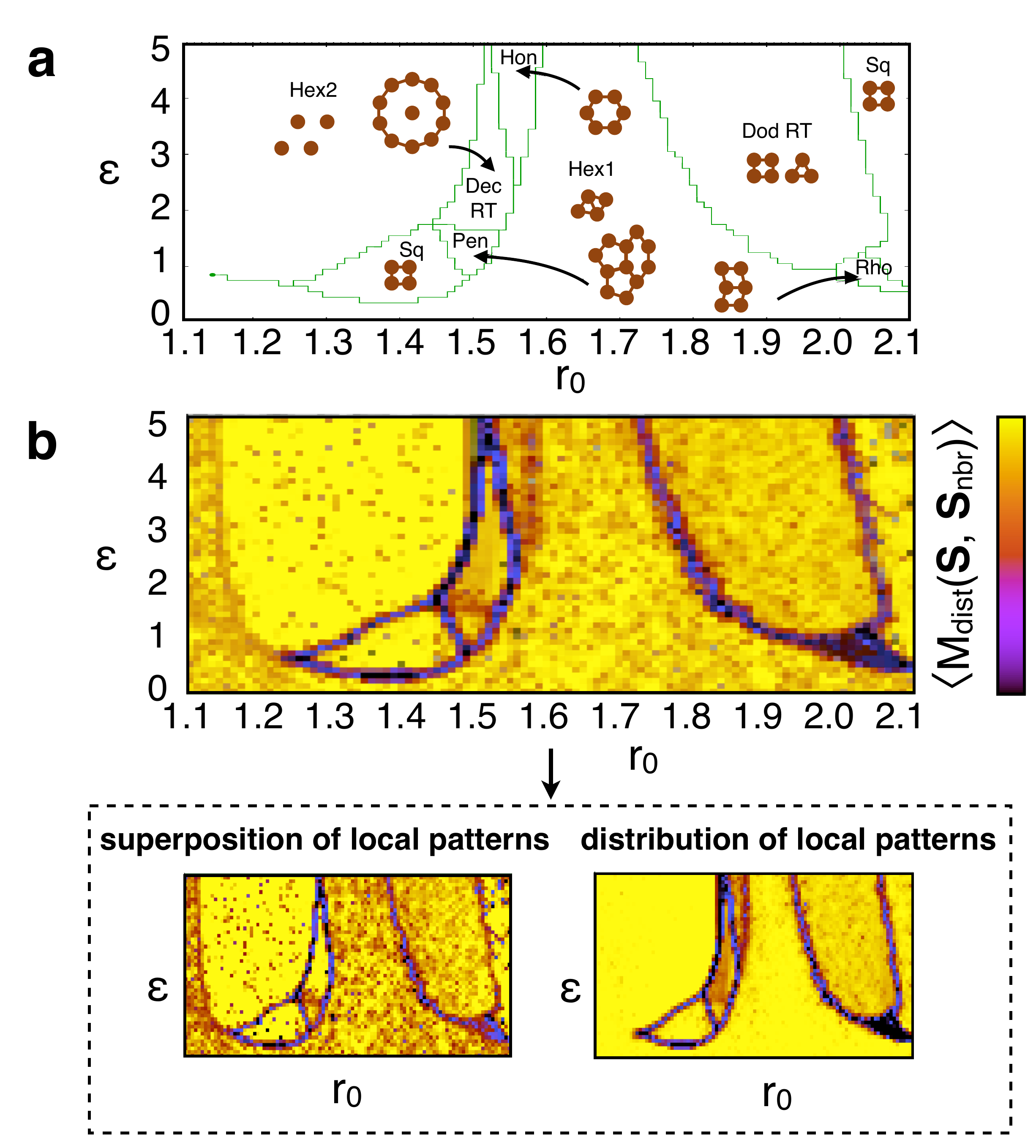}
\caption[Structural Phase Diagram for the 2d Lennard-Jones Gauss System]{Structural phase diagram for the 2d Lennard-Jones Gauss system~\cite{engel07}.  (\textit{a}) Structural phase diagram created by visual inspection~\cite{engel07}.  (\textit{b}) Structural phase diagram created by shape matching.  Each pixel in parameter space is given an intensity based on the average match with structures for neighboring points.}
\label{fig:autophase}
\end{center}
\end{figure}

The general ideas underlying the correlation functions outlined above can be applied to more abstract problems as well.  For example, rather than creating correlation functions in space and time, we can create correlation functions in parameter space.  Consider the problem of automatically generating the structural phase diagram for the 2D Lennard-Jones-Gauss (LJG) system~\cite{engel07} shown in Fig.~\ref{fig:autophase}a from a collection of simulation snapshots for each statepoint.  The phase diagram was generated by visual inspection of over 5000 statepoints~\cite{engel07}.  We can automate the creation of this phase diagram by using an idea similar to the $\textbf{q}_6 \cdot \textbf{q}_6$ scheme outlined above.  In this case, rather than finding structural correlations between neighboring particles in real-space, we can calculate correlations between neighboring snapshots in parameter space.  For each point on the structural phase diagram, we compute $I(i) = \sum_j M(\textbf{S}_i,\textbf{S}_j)$, where ``$i$'' and ``$j$'' are neighbors in parameter space, and $\textbf{S}$ is a global shape descriptor.  Here, we take the global descriptor as a combination of a global superposition descriptor, indexed by a shape histogram with $n_\theta = 20$, $n_r = 1$, and a global probability distributions descriptor based on local Fourier descriptors with frequency range $\ell=6, 7, ... 11$: $\textbf{S}^\mathrm{global} = \left< \textbf{S}^\mathrm{H2}_{global}, P(\textbf{S}^\mathrm{F2}_{local}) \right>$. Points in stable regions of parameter space match well with neighboring points and have a high value of $I(i)$, whereas transitional points match poorly and have a low value of $I(i)$.  The algorithm for creating a visual map of the structural phase boundaries for a parameter space is given in pseudocode below:
\begin{quote}
\begin{mylisting}
\begin{verbatim}
for each point i in parameter space 
   call compute_shape_descriptor(S[i])
end for
 
for each point i in parameter space 
    set pixel_intensity[i] = 0
    for each neighboring point j 
        set pixel_intensity[i] += M(S[i], S[j])
    end for
 end for
\end{verbatim}
\end{mylisting}
\end{quote}

The transitional points map out structural phase boundaries that look very similar to the diagram created by visual inspection.  Notice that the global superposition descriptor detects no difference between the hexagonal and honeycomb crystals, since both structures are six-fold symmetric, and thus yield equivalent combined shape histograms.  Additionally, the superposition descriptor picks up a slight artificial ``boundary'' within the hexagonal region near $r_0 \sim 1.15$, where the hexagonal crystal begins to form multiple grains rather than a single crystal, resulting in different shape histograms.  The probability distributions descriptor, on the other hand, gives no distinction between the pentagonal and decagonal phases, since they are nearly identical locally, differing only in long range ordering.  Overall, both sets of global information taken in combination are necessary to correctly find all of the phase boundaries for this particular application.  However, in many cases, the space of structures is sufficiently non-degenenate to be described by a single method.


\subsection{Heat Maps and Grouping}
\begin{figure}
\begin{center}
\includegraphics[width=1.0\columnwidth]{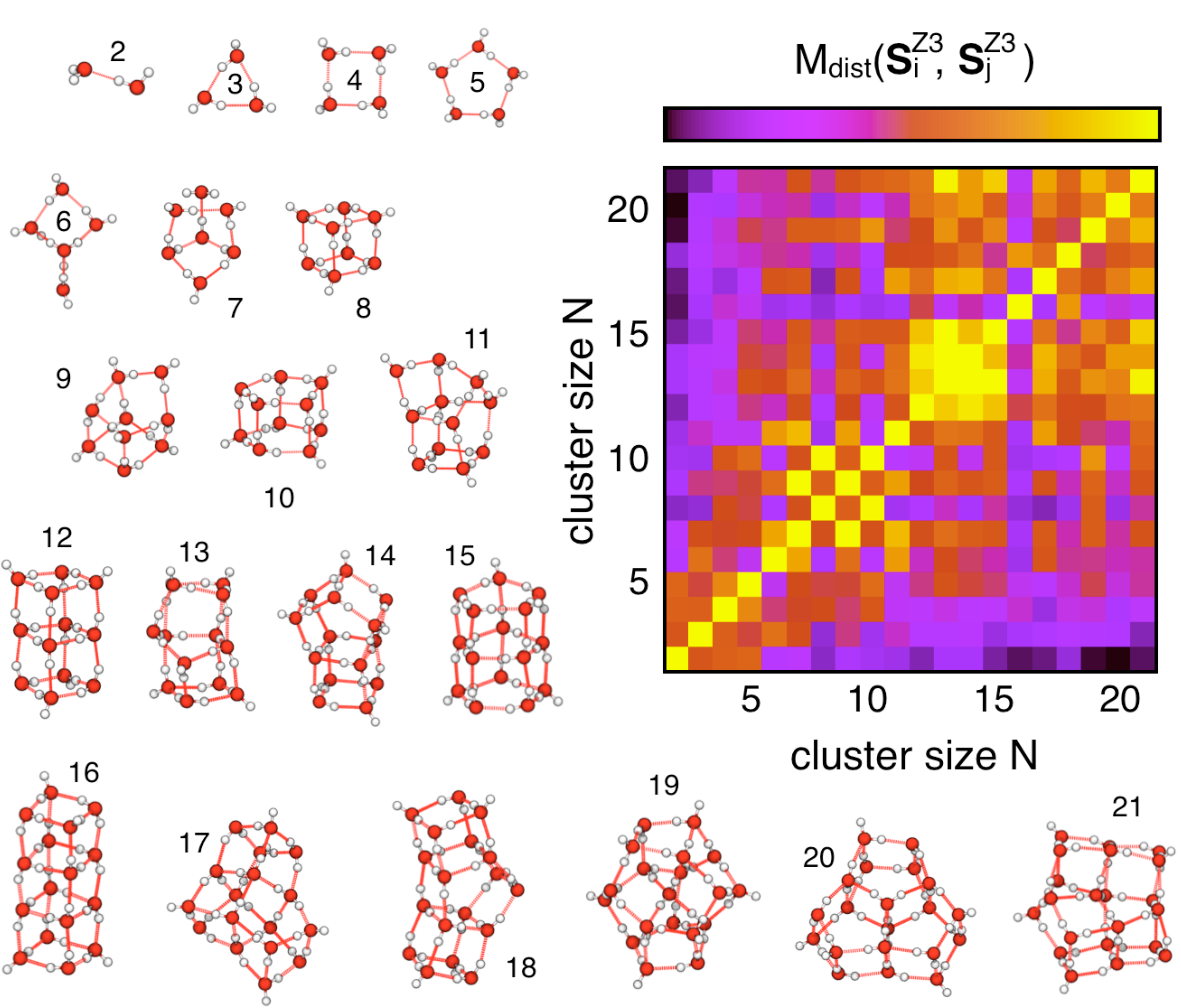}
\caption[Grouping and Classifying Structures Based on Shape Similarity]{Grouping and classifying structures based on shape similarity.  The plot shows a similarity matrix for energy-minimized TIP4P water clusters~\cite{water}.  The matrix simultaneously shows the pairwise matching values for all clusters. }
\label{fig:water}
\end{center}
\end{figure}

Another common application of shape matching techniques is to the problem of visually grouping or classifying similar structures~\cite{princetonshape}.  Grouping objects based on shape similarity has also been applied recently to macromolecules and proteins~\cite{mak, yeh}.  Grouping can be accomplished by plotting the matrix of pairwise matching values known as a  ``similarity matrix'' or ``heat map.''   The algorithm for computing a similarity matrix is given in pseudocode below:
\begin{quote}
\begin{mylisting}
\begin{verbatim}
for i in list_of_local_structures
   call compute_shape_descriptor(S_i)
   for j in list_of_local_structures
       call compute_shape_descriptor(S_j)
       set matrix[i,j] =  M(S_i, S_j)
end for
return matrix
\end{verbatim}
\end{mylisting}
\end{quote}
Objects can be grouped or classified based on features of the plot.  As an example, consider the TIP4P water clusters~\cite{water} shown in Fig.~\ref{fig:water}.  The matrix shows the match values obtained for minimum energy clusters of sizes $N=2-21$, which are available from the Cambridge Cluster Database~\cite{ccd}.   The clusters are indexed using rotation-invariant Zernike descriptors with frequency parameters $\ell=4, 5, .. 12$. 

The patterns displayed in the heat map require some interpretation. The high correlation along the diagonal $i=j$ is common to all heat maps, and simply indicates that structures match perfectly with themselves.  The region $N=12-15$ displays a bright box, which indicates a group of structures that all match well with one another.  The region $N=7-10$ displays a checkerboard pattern, which indicates that every-other cluster matches well.  Cluster $N=16$, due to its unique non-compact nature, matches poorly with all other clusters, as indicated by the dark (purple) cross at $N=16$. 
In addition to grouping and classifying objects, heat maps can be used to visually indicate convergence with respect to a changing parameter.  Multiple heat maps based on different descriptors can be constructed for the same set of structures to show the similarities on different levels of ordering.  Although we consider local clusters for our example, heat maps can be applied to any structures that can be indexed by shape descriptors, including global structures.

\section{Future Outlook}

In summary, we have introduced a shape matching framework for creating new structural metrics for complex patterns, such as those encountered in self-assembly.  All of the methods and examples outlined here are accessible online through our C/C++ shape matching library~\cite{smwebsite}.  Although our examples and discussions here are geared towards self-assembly and condensed matter physics, the general ideas underlying the shape matching framework are widely applicable to systems with complex structures, such as those encountered in computational biology~\cite{keys10-HO}.  In the future, new shape descriptors and algorithms can be added to the framework to expand its scope to different classes of structures and problems.  

The example applications and shape descriptors that we have presented here represent only a small subset of the vast range of possibilities yet to be explored.  One obvious area for future study is to test the applicability of the wealth of shape descriptors from the shape matching literature to particle systems.  Another promising area to explore is the creation of new abstract order parameters and correlation functions, such as the phase space correlation function of Fig.~\ref{fig:autophase}.  This type of application may represent one of the most important uses for shape matching moving forward;  replacing the human element with a computer algorithm to explore parameter space has the potential to greatly expedite self-assembly research.


\textbf{Acknowledgements:} ASK was partially supported by a grant from the U.S. Department of Education (GAANN Grant No. P200A070538). SCG and ASK received partial support from the U.S. National Science Foundation (Grant Nos. DUE-0532831 and CHE-0626305). SCG and CRI received support from the U.S. Department of Energy, Office of Basic Energy Sciences, Division of Materials Sciences and Engineering under Award \#  DE-FG02-02ER46000.  CRI also acknowledges the University of Michigan Rackham Predoctoral Fellowship program. We thank G. van Anders and T.D. Nguyen for helpful comments on the manuscript. Thanks also to T.D. Nguyen, M. Engel, A. Haji-Akbari, E.P. Jankowski, C. Phillips, D. Ortiz, A. Santos, I. Pons, and C. Singh for providing example data, not all of which could be used here.

\bibliography{References}
\bibliographystyle{ieeetr}

\end{document}